\documentclass[aps,rmp,twocolumn,showpacs,groupedaddress,longbibliography,preprintnumbers]{revtex4-1}
\pdfoutput=1
\usepackage{graphicx} 
\usepackage{color}

\usepackage{amsmath}  
\usepackage{amssymb}
\usepackage{ulem}
  \newcounter{runningcount}
\definecolor{IITred}{rgb}{0.5,0.05,0.05}
\definecolor{Dgreen}{RGB}{0,96,0}
\definecolor{Dcyan}{cmyk}{.96,0,0,.4}
\definecolor{IITblue}{rgb}{0.05,0.05,0.8}

\usepackage{hyperref}
\hypersetup{
    colorlinks=true,
    linkcolor=blue,
    filecolor=magenta,      
    urlcolor=Dcyan,
    citecolor=IITblue,
      }

\newcommand{\y}{\text{ yr}}
\newcommand{\gy}{\text{ Gyr}}
\newcommand{\s}{\text{ s}}
 \newcommand{\onetev}{1-TeV scale}
    \newcommand{\ev}{\hbox{ eV}}
    
    \newcommand{\mev}{\hbox{ MeV}}
    \newcommand{\gev}{\hbox{ GeV}}
      \newcommand{\gevcc}{\hbox{ GeV}/c^2}
  \newcommand{\tev}{\hbox{ TeV}}
    \newcommand{\m}{\hbox{ m}}
    \newcommand{\cm}{\hbox{ cm}}

    \newcommand{\K}{\hbox{ K}}
    
\newcommand{\half}{\ensuremath{\scriptstyle \frac{1}{2}}}     

\newcommand{\jpsi}{\ensuremath{J\!/\!\psi}}

\newcommand{\smgg}{\ensuremath{\mathrm{SU(3)_c} \otimes \mathrm{SU(2)_L} \otimes \mathrm{U(1)}_Y}}
\newcommand{\ewgg}{\ensuremath{\mathrm{SU(2)_L} \otimes \mathrm{U(1)}_Y}}
\newcommand{\emgg}{\ensuremath{\mathrm{U(1)}_{\mathrm{em}}}}
\newcommand{\legg}{\ensuremath{\mathrm{SU(3)_c}  \otimes \mathrm{U(1)}_{\mathrm{em}}}}

\def\abs#1{\left| #1\right|}
\def\vev#1{\langle #1\rangle_0}

\usepackage{breakurl}

\newcommand{\alphas}{\ensuremath{\alpha_{\mathrm{s}}}}
\newcommand{\orcid}[1]{\thanks{\href{http://orcid.org/#1}{ORCID: #1}}}

\begin{document}
\title{Perspectives and Questions: \\ {Toward an Expansive Agenda for Particle Physics}}
\preprint{\sffamily FERMILAB--PUB--25/0689--T}
\author{Chris Quigg}
\email[]{quigg@fnal.gov}
\orcid{0000-0002-2728-2445}
\affiliation{Fermi National Accelerator Laboratory \\ {P.O.\ Box 500, Batavia Illinois 60510 USA}}
\date{\today} 

\vspace*{6pt}
\begin{abstract}
Global celebration greeted the 2012 discovery at CERN's Large Hadron Collider of a particle that matches the textbook description of the Higgs boson. That achievement validated a remarkable chain of theoretical reasoning that combined the  prescriptive notion of  electroweak gauge symmetry with a simple, but \emph{ad hoc}, embodiment of spontaneous symmetry breaking. It was made possible by generational triumphs of accelerator art and experimental technique,  and by human resourcefulness and collaboration on a global scale, all sustained by the enlightened support of many governments and institutions.  \\

Some  imagine that, once the keystone of the standard model of particle physics has been set, our subject is over. Others worry that we may be at an impasse because no comparable wonders have appeared, leaving us without well-defined clues to a more complete paradigm. I am neither so readily satisfied nor so {easily} discouraged: we have so much more to learn! This essay surveys many questions  that, taken together, constitute an inspiring  array  of opportunities to enhance our understanding of the physical world.
  \end{abstract}
\maketitle

\tableofcontents

\section{Prologue\label{sec:invite}}

 \subsection{Anticipations of a New World \label{subsec:ExoticFruits}}

Before experiments\footnote{The LHC detectors~\cite{Froidevaux2006} and upgrades planned for the (high-luminosity) HL-LHC~\cite{Campana2016} represent the greatest complexity and performance yet achieved. The developments in computing needed to keep pace are detailed in~\cite{Elvira:2022wyn}.} began at the Large Hadron Collider, we  were confident that a thorough exploration of the \onetev\ would  reveal the mechanism of electroweak symmetry breaking. We had reason to suspect that in this same energy range we might discover (WIMP, weakly interacting massive particle) dark matter candidates, superpartners, or evidence for new strong dynamics related to dynamical electroweak symmetry breaking. And why not new constituents or new gauge bosons, representing new forces of Nature?

Looking ahead, I was encouraged by these words of Cecil Frank Powell, a pioneer of the photographic emulsion method that revealed the pion~\cite{Powell:frag}:
\begin{quote}{\sffamily 
 ``When [the emulsions exposed on the Pic du Midi]  were recovered and developed in Bristol it was immediately apparent that a whole new
world had been revealed. \ldots  It was as if, suddenly, we had broken into a walled orchard, where protected trees had
flourished and all kinds of exotic fruits had ripened in great profusion.''}
\end{quote}
Some optimists asserted that a mere eight minutes of LHC experimentation would establish the existence of supersymmetric particles. That sort of exuberance---imprudent at the time---{invites ridicule} in retrospect.
In the event, we have reaped ``only'' the epochal discovery of the 125-GeV Higgs boson~\cite{ATLAS:2012yve,CMS:2012qbp}\footnote{On a recent visit to the Museo Galileo in Florence, I came upon these effusive words from one Ludovico (Antonio) Muratori (1672--1750): ``God has truly reserved to our own day the discovery of an incredibly amazing phenomenon. I am referring to electricity.''
Should we not in turn temper our laudable impatience a little and  celebrate the fact that Nature has reserved to our day the discovery of a most wondrous phenomenon, the agent of electroweak symmetry breaking?}.

While it is true that we have found no other new terms in the Lagrangian of the universe, each negative search offers fresh guidance for theory. And if we broaden our view of discovery just a little, we find that the LHC has indeed revealed exotic fruits in great profusion, in the form of seventy-nine new hadrons~\cite{*[{}][{. For new hadrons observed at other colliders, see~\cite{QWGExotics}.}]PKopp}, a good number expressing body plans\footnote{This  term from morphology and comparative anatomy encompasses both tabulating constituents and identifying substructures.}  beyond the classic meson $(q\bar{q})$ and baryon $(qqq)$ configurations. 

Progress in particle physics is by no means confined to experimentation at the highest energies. Our subject is continually nourished by news from exquisitely precise measurements, experiments using found beams from natural sources, and a widening spectrum of astrophysical and cosmological observations.
Synthesizing all this information is a fascinating challenge for contemporary research.

In this essay, I survey the frontiers of particle physics by posing a multitude of questions. The general message is that the openings for progress are many and broadly distributed. Each question could merit an essay of its own, but I have adopted the format of lists, not fully documenting the questions. The references are schematic; I trust that a curious reader will be able to enrich them\footnote{Two essential resources are the Snowmass 2021 Report~\cite{Butler:2023glv} and the \emph{Physics Briefing Book: Input for the 2026 update of the European Strategy for Particle Physics}~\cite{deBlas:2944678}.}. 

 My goals are to encourage colleagues (myself included) to think broadly and originally about our common future. I hope that readers will  consider the opportunities I catalogue, to make their own lists---and perhaps to order them and set personal and community priorities, conscious of the advantages of diversity and of scale diversity in experimental undertakings.
 It would please me if the list that follows served as incentive for a graduate seminar. An open-ended  assignment could be to choose a question, flesh it out, investigate, report, refine the question, and even act on it.

 \subsection{Questions Our Forebears  Pondered \label{subsec:NALreport}}
 A revealing way to assess the progress in particle physics over the past six decades is to consider the Problems of High-Energy Physics cited in the (Fermi) National Accelerator Laboratory Design Report~\cite{Cole:1968wx}:
\begin{quotation}{\sffamily
Which, if any, of the particles that have so far been discovered, is, in fact, elementary, and is there any validity in the concept of ``elementary'' particles? 

What new particles can be made at energies that have not yet been reached? Is there some set of building blocks that is still more fundamental than the neutron and the proton? 

Is there a law that correctly predicts the existence and nature of all the particles, and if so, what is that law? 

Will the characteristics of some of the very short-lived particles appear to be different when they are produced at such higher velocities that they no longer spend their entire lives within the strong influence of the particle from which they are produced? 

Do new symmetries appear or old ones disappear for high momentum-transfer events? 

What is the connection, if any, of electromagnetism and strong interactions? 

Do the laws of electromagnetic radiation, which are now known to hold over an enormous range of lengths and frequencies, continue to hold in the wavelength domain characteristic of the subnuclear particles? 

What is the connection between the weak interaction that is associated with the massless neutrino and the strong one that acts between neutron and proton? 

Is there some new particle underlying the action of the "weak" forces, just as, in the case of the nuclear force, there are mesons, and, in the case of the electromagnetic force, there are photons? If there is not, why not? 

In more technical terms: Is local field theory valid? A failure in locality may imply a failure in our concept of space. What are the fields relevant to a correct local field theory? What are the form factors of the particles? What exactly is the explanation of the electromagnetic mass difference? Do ``weak'' interactions become strong at sufficiently small distances? Is the Pomeranchuk theorem\footnote{Let $a$ and $b$ label hadron species. The inference that differences of the  $ab, \bar{a}b, a\bar{b}, \text{and } \bar{a}\bar{b}$ total cross sections should vanish at high energies is due to \cite{Pomeranchuk:1958ged}.} true? Do the total cross sections become constant at high energy? Will new symmetries appear, or old ones disappear, at higher energy? 

}
\end{quotation}
Young physicists in particular may be a bit astonished by how little our colleagues knew in 1968 of what we consider textbook material. But consider how much more they knew than the pioneers of a half century before, and be prepared to be chastened by how quaint and incomplete our current knowledge will seem a few decades hence! While some of the questions are challenging to decrypt, overall they exhibit a great deal of insight into the kind of issues that might matter. Indeed, several   remain relevant in our time.
\subsection{Context\label{subsec:context}}
\begin{figure}[tbh]
\centerline{\includegraphics[width=0.4\textwidth]{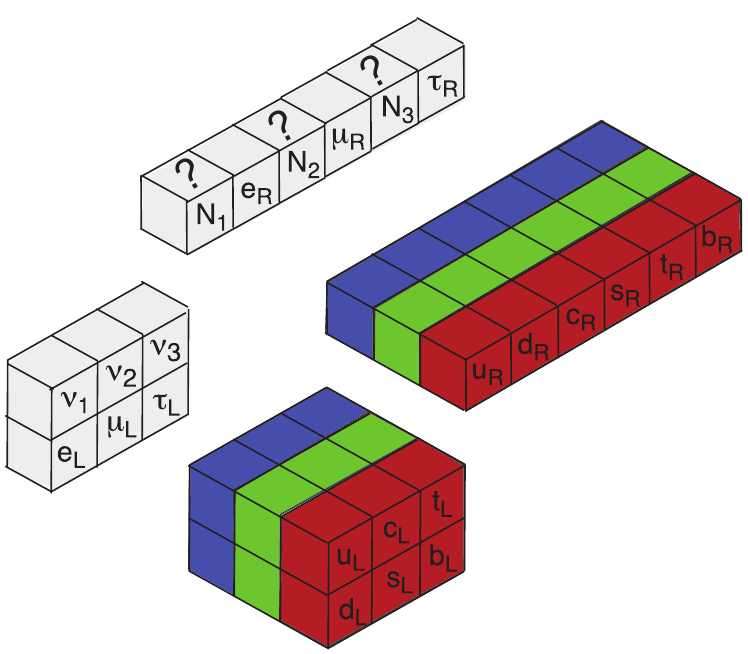}}
\caption{A schematic representation of the quarks and leptons, the fundamental constituents of the standard model of particle physics. The color-triplet quarks are painted red, green, and blue; the left-handed weak-isospin doublets are indicated by the stacked pairs. Interactions are governed by \smgg\ gauge symmetry, spontaneously broken to \legg.  \label{fig:SMcartoon}}
\end{figure}

Let us set our point of departure: the evolving standard model of particle physics circa 2025. We have established a set of building blocks, spin-\half\ constituents that appear pointlike at the current limit of our resolution, $r \lesssim 10^{-18}\m$. These are arrayed in Figure~\ref{fig:SMcartoon} to exhibit their family relationships.

Six quarks (up, down, charm, strange, top, and bottom) are color triplets that experience the strong interaction described by quantum chromodynamics (QCD), generated by the color gauge group $\mathrm{SU(3)_c}$. The left-handed quarks experience the charged-current weak interaction; they form weak-isospin doublets that are roughly aligned with the mass eigenstates: $(u, d), (c, s), (t, b)$. On current evidence, the right-handed quarks are weak-isospin singlets.

Six color-singlet left-handed leptons ($\nu_e, e, \nu_\mu, \mu, \nu_\tau, \tau$) form weak-isospin doublets in which the flavor eigenstates are perfectly aligned. In the figure, I have labeled the neutrinos as mass eigenstates $(\nu_1, \nu_2, \nu_3$), to match the labeling of the quarks. We infer the likely existence of right-handed neutrinos (here labeled $N_1, N_2, N_3$) from the discovery that neutrinos have (very tiny) masses. The \textsf{?} signal that we have not yet established that the right-handed neutrinos exist, nor characterized them in detail.

In addition to the color and left-handed weak isospin family symmetries, there is a weak-hypercharge phase symmetry, so that the interactions that rule the nanonanoworld\footnote{The distance scale around $10^{-18}\m$ might properly be called the attoworld, I use \emph{nanonano} to emphasize how far we have come from the atomic scale.} are derived from a $\smgg$ gauge symmetry, that is hidden, or spontaneously broken, to \legg. The placement of quarks and leptons in Figure~\ref{fig:SMcartoon} makes it visually tempting to join the quark and lepton multiplets together to form extended quark--lepton families, with lepton number as a fourth color~\cite{Pati:1974yy}.

To build a theory out of these observations, we lay out a small set of Guiding Principles. We take Nature to be reliable, not capricious. We adopt Poincar\'e-invariant Relativistic Quantum Field Theory, embodying local causality and unitarity as a productive mathematical framework, disciplined by the constraints of symmetries and conservations laws (via Noether's theorems); and the notion of hidden symmetries. 
 A generation ago, Renormalizability would have been taken as a \emph{sine qua non.} Now we are comfortable with the notion of {effective field theories}~\cite{Weinberg:2021exr}, in which we identify the important degrees of freedom for each scale of energy or resolution.

To this general framework, I would add these working hypotheses:\\
$\Box$ Minkowski spacetime (except where gravity is strong),\\
$\Box$ constancy in time of the fundamental constants such as $\hbar, c, G_\mathrm{N}, m_e, \alpha, \ldots$, \\
$\Box$ pointlike quark and lepton constituents with sharply defined masses,\\
$\Box$ gauge symmetry as the typical origin of interactions.


The product $\mathcal{CPT}$ of the discrete
symmetries, charge conjugation $\mathcal{C}$, parity $\mathcal{P}$, and time reversal $\mathcal{T}$, is an exact symmetry of any local and Lorentz-invariant quantum field theory
with a positive-definite Hermitian Hamiltonian that preserves microcausality~\cite{*[{For a comprehensive treatment, see }][]PCTSpin}.
All observations to date indicate that the strong and electromagnetic interactions respect charge conjugation, parity, and time reversal invariance separately, but  the weak interactions do not\footnote{For a review of tests of conservation laws for discrete spacetime (and other symmetries), see A. Pich and M. Ramsey-Musolf, ``Tests of Conservation Laws,'' in \cite{ParticleDataGroup:2024cfk}.}.


 \subsection{How to Progress?\label{subsec:howto}}
 To advance our understanding, we will need to explore  regions of the unknown and to probe  unanswered questions. It is the role of all of us, and an obligation of theorists in particular, to try to divine where the secrets are hidden.
We must seek out soft spots in our current understanding,
especially where the stories we tell are
{\emph{unprincipled,}} by which I mean not (yet) founded on sound principles.

Think, for example of \emph{Supersymmetry,} a beautiful and rich idea. As soon as we try to apply it to our world, we must add a new element, such as $R$-parity, to ensure a measure of proton stability. Then, if we want to put supersymmetry forward  as a solution to naturalness, we must solve the so-called $\mu$ problem, to make it relevant on the \onetev. Generically, supersymmetric theories imply flavor-changing neutral currents that are not observed, so we must find a mechanism that controls those. These issues do not mean that supersymmetry cannot be a valid extension of the standard model, but that we have not yet established principles that will make it so.

Similarly, \emph{Big-Bang Cosmology} is indicated by a wealth of observations. Once we consider the idea of a hot, dense early universe, we must add a concept (usually taken to be inflation) to account for the exquisite flatness and uniformity of the universe and other features. As we learn more of the composition of the universe over time, it seems that we must add dark matter and dark energy, and perhaps other ingredients. All of which is to say that our impressive ``standard cosmology'' is still a work in progress---which is not a bad thing.

And let us confess that we do not know what, if anything, establishes the particle content and the gauge groups of the standard model of particle physics.

 The first set of questions explores our Guiding Principles and Working Hypotheses, for they must not pass unchallenged.
 \begin{enumerate}
\setcounter{enumi}{\value{runningcount}}
\item Is Lorentz invariance exact\footnote{For compendia of tests of Lorentz invariance, see~\cite{Kostelecky:2008ts} and \cite{Liberati:2013xla}
}?
\item Are Nature's laws the same at all times and places (accessible to us)?
\item What is the domain of validity of local quantum field theory? 
\item What would it mean for causality to be violated? 
\item Is $\mathcal{CPT}$ a good symmetry?
\item Are there novel sources of $\mathcal{C, P, T,\mathrm{ and }~CP}$  violation? Can we find evidence for $\mathcal{P}$ and $\mathcal{CP}$-violating permanent electric dipole moments? 
\item Do quarks and leptons show signs of compositeness?  Are they made of more elementary constituents? \label{item:comp}
\item Can we find evidence for states with indefinite mass (beyond ordinary unstable resonances), such as unparticles or mass-varying neutrinos?
\item Are there supplemental spacetime dimensions?
\item Do any of the fundamental constants vary with time?
\setcounter{runningcount}{\value{enumi}}
\end{enumerate}

%

\section{Beyond the Higgs-Boson Discovery}
The discovery of the Higgs boson by the ATLAS~\cite{ATLAS:2012yve} and CMS~\cite{CMS:2012qbp} Collaborations working at the LHC is a landmark for our understanding of Nature and a remarkable achievement of many people---especially the experimenters and accelerator builders whose extended effort made the discovery happen\footnote{My before-and-after articles, \cite{Quigg:2009vq} and \cite{Quigg:2015cfa}, provide some intellectual history, reviews of theoretical contributions, and context.}. We can say of the new, unstable particle with mass $M_H = 125.20 \pm 0.11\gev$ \cite{ParticleDataGroup:2024cfk} that the evidence is developing as it would for a textbook Higgs boson of the standard electroweak theory~\cite{ATLAS:2022vkf,CMS:2022dwd,Salam:2022izo,ATLAS:2024lyh,CMS:2025jwz}. Its observed decays into $W^+W^-$ and $ZZ$ implicate $H(125)$ as an agent of electroweak symmetry breaking\footnote{In a \emph{Gedanken} world lacking a Higgs mechanism, Quantum Chromodynamics breaks the electroweak symmetry $\ewgg\to\emgg$, but induces only a tiny $M_W \approx 28{\mev}$~\cite{Quigg:2009xr}.}. The $W$-boson mass (through the factor $1/M_W^4$) controls $\beta$-decay rates and energy production in the Sun.  The putative Higgs boson decays to $\gamma\gamma$ at approximately the expected rate. It is dominantly spin-parity $J^P=0^+$. Evidence for the $Ht\bar{t}$ coupling from the dominant production mechanism, gluon fusion through a top-quark loop, and from the  observation of $t\bar{t}H$ production, imply that the Higgs field plays a role in the generation of fermion masses. That implication is supported by the observation of decays into $\tau^+\tau^-$, $b\bar{b}$, and lately  $\mu^+\mu^-$, at close to the expected rates. 

The LHC experiments are sensitive to the gluon-fusion and vector-boson--fusion production mechanisms, associated production of a Higgs boson plus an electroweak gauge boson, and $Ht\bar{t}$ or $Htq$ production. At the current precision, the observed yields, which measure production cross section times branching ratios, are in line with standard-model expectations. We have found no evidence yet for charged or neutral companions to $H(125)$, and no suggestion of new strong dynamics. Although there is no direct measurement of the Higgs-boson width, deft analysis of interference effects  plausibly yields a value of the width  close to the standard-model expectation.

This already impressive dossier indicates that the prospecting (or search-and-discovery) phase is over, and that we have advanced to a painstaking forensic investigation. What remains to be learned? Ever more quantitative studies within the framework of effective field theory will play a leading role~\cite{Dawson:2018dcd}.  It is important to test whether the Higgs field is also a giver of mass to the lighter quarks and to the electron. Finding all these decay modes at the expected rates is an important branch point for theories of the fermion masses: it could rule against some pictures in which the source of light-fermion masses is quantum effects tied to the heavy fermions. Verifying the standard-model $H\to\mu^+{\mu^-}$ coupling  raises interest in a $\mu^+\mu^- \to H$ factory. Some studies indicate that a muon collider's beam-momentum spread might be fine enough to permit mapping out the $H$ line shape, which would be a most impressive feat. 

Establishing the origin of the electron's mass occupies a special place in our quest to understand the nature of matter. The electron mass sets the scale of atomic energy levels ($\propto m_ec^2\alpha^2$ in a Coulomb potential). It also governs the Bohr radius, and so the size and integrity of atoms, prerequisites to valence bonding. If I ruled the world, I would award the Nobel Prize in Chemistry to whoever shows that the Higgs field is responsible. This is no easy task: the $H\to e^+e^-$ branching fraction  is only about five parts per billion. FCC-ee enthusiasts express hope that the formation reaction $e^+e^- \to H$ might be in reach, but we are far from knowing that it could be done. 

The mass difference between up and down quarks determines whether protons or neutrons are stable against $\beta$-decay. The role of the Higgs field in setting these masses is not yet established.

The LHC experiments access several production channels and many decay modes; they will provide much additional precise information. But it would be advantageous to have a second look through the reaction $e^+e^- \to HZ$. If a high-luminosity $e^+e^-$ Higgs factory were to drop out of the sky tomorrow, the line of users would be very long. Although that marvelous event will not happen, several ambitious proposals are in view. Their particular assets include the ability to determine absolute branching fractions and to measure directly the total width of the Higgs boson, and graceful access to decay channels such as $H \to c\bar{c}$. Important complementary information could be gleaned in other operating modes: Tera-$Z$ for both precision tests and discovery, a $WW$ threshold scan, and studies at $t\bar{t}$ threshold. A more comprehensive set of precise measurements, from the LHC, HL-LHC, and future machines, will enable us to make ever more incisive tests of the standard model as a quantum field theory, and to reflect on the implications of $M_H \approx 125\gev$.

I close this section with a list of questions about electroweak symmetry breaking and the Higgs sector that we must answer to approach a final verdict about how closely $H(125)$ matches the textbook Higgs boson. 
\begin{enumerate}
\setcounter{enumi}{\value{runningcount}}
\item Is $H(125)$ the only member of its clan? Might there be others---charged or neutral---at higher or lower masses?
\item Does $H(125)$ fully account for electroweak symmetry breaking? Does it match standard-model branching fractions to gauge bosons? In greater depth, are the absolute couplings to $W$ and $Z$ as expected in the standard model?
\item Is the Higgs field the only source of charged fermion masses? Are all the fermion couplings proportional to fermion masses?
\item Can we find evidence for flavor-changing decays of $H(125)$?
\item What can we learn from the rare decays $H\to\Upsilon\,\gamma$ and $H\to\jpsi\,\gamma$, and the comparison to the $H\to b\bar{b}, c\bar{c}$ decay rates?
\item What accounts for the immense range of fermion masses?
\item How can we detect  $H \to e^+e^-$?
\item What role does the Higgs field play in generating neutrino masses? 
\item Can we establish or tightly constrain decays to new particles? Does $H(125)$ act as a portal to hidden (dark or subliminal) sectors? 
\item How convincingly can we measure $\Gamma_H$ and compare with theory?
\item Are all production rates as expected? Will we uncover any surprising sources of $H(125)$?
\item Do loop-induced decays such as $gg, \gamma\gamma, \gamma Z$ occur at standard-model rates?
\item Can we find any sign of new strong dynamics or (partial) compositeness?
\item Can we establish the $H\!H\!H$ trilinear self-coupling?
\item How well can we test the notion that $H$ regulates Higgs--Goldstone scattering, i.e., tames the high-energy behavior of longitudinal $WW$ scattering~\cite{Lee:1977eg}?
\item What is the order of the electroweak phase transition?
\item Does the electroweak vacuum seem stable, or suggest a new physics scale? 
\item What conditions must we create to restore \ewgg\ symmetry at high energies? How will we know?
\setcounter{runningcount}{\value{enumi}}
\end{enumerate}
The last six entries call for sensitive studies at high energies---almost certainly higher than we have available at the $\sqrt{s} \approx 14\tev$ Large Hadron Collider.

\section{More Insights on the TeV Scale \& Beyond?}
Before experiments began at the LHC, there was much informed speculation---but no guarantees---about what might be found, beyond the keys to electroweak symmetry breaking. The targets included supersymmetry and technicolor, either of which might have served as a once-and-done solution to enforcing the large hierarchy between the electroweak scale and the unification scale or Planck scale. Did these two potential solutions lead us to view the hierarchy problem and the discipline of ``naturalness'' too simplistically~\cite{Bardeen:1995kv,WILSON20053,Dine:2015xga,Giudice:2017pzm}?

Our optimism was also encouraged by the observation that a dark-matter candidate in the form of a weakly interacting massive particle would naturally reproduce (what we take to be) the observed relic density, if the WIMP mass lay in the range of a few hundred GeV. We cannot prove that an apparently stable particle produced in the collider environment has a cosmological lifetime, but if we were to produce a candidate we could explore its properties in much greater detail than we imagine doing in direct- or indirect-detection experiments. If our reading of the evidence for dark matter is correct, we will need to assemble evidence from all the experimental approaches. {We will examine evidence from the state of the universe in \S\ref{sec:Astro/Cosmo}.}

No direct sign of new physics beyond the standard model has come to light in laboratory experiments, but searches must continue at the LHC and beyond. The first hints may come from precision measurements---think of the (perhaps resolved) $(g - 2)_\mu$ anomaly or the fading hints of lepton nonuniversality---rather than the direct observation of new phenomena. Studies of rare processes also give us virtual access to energy scales far beyond what we can reach directly today, or perhaps ever.  And there is the nagging headache that the vacuum expectation value of the Higgs field, $\vev{H}$, if taken at face value, contributes a staggeringly large energy density throughout the universe: the equivalent of one Jupiter mass inside a typical person.

These considerations invite further questions.
\begin{enumerate}
\setcounter{enumi}{\value{runningcount}}
\item Are there new forces of a novel kind?
\item Can we find evidence of a dark matter candidate (or more than one)?
\item Why is empty space so nearly massless? What is the resolution of the vacuum energy problem? 
\item Will ``missing energy'' events or Ka\l uza--Klein excitations of the graviton signal the existence of spacetime dimensions beyond the familiar $3+1$?
\item Can we find clues to the origin of electroweak symmetry breaking? Is there a dynamical origin to the ``Higgs potential?''
\item What separates the electroweak scale from higher scales?
\item Might we find indirect evidence for a new family of strongly interacting particles, such as those that are present in supersymmetric extensions of the standard model, by seeing a change in the evolution of the strong coupling ``constant,'' $1/\alphas$, at a higher-energy LHC or a ``100-TeV'' collider?
\item Might new phenomena appear in collider experiments at macroscopic scales (not close to a primary production vertex)?
\item How can we constrain---or provide evidence for---light dark-matter particles or other denizens of the dark in high-energy colliders or beam-dump experiments?
\item Does the gluon have heavy partners, indicating that QCD is part of a gauge structure richer than $\mathrm{SU(3)_c}$?
\item How can we mount telling searches for magnetic monopoles~\cite{MoEDAL:2014ttp} or sphalerons~\cite{Ellis:2016ast,Grefsrud_2024} or black holes or other exotics?
\setcounter{runningcount}{\value{enumi}}
\end{enumerate}

\section{Flavor I: the Problem of Identity}
For the issue of electroweak symmetry breaking,  the central questions were clearly articulated for many years and we identified the 1-TeV scale as the promised land for finding answers. In contrast, we do not have a clear view of how to approach the diverse character of the constituents of matter---the quarks and leptons. To be sure, we have challenged the Cabibbo--Kobayashi--Maskawa (quark-mixing matrix) paradigm and found it an extraordinarily reliable framework in the hadron sector. Neutrino oscillations, discussed in the following Section~\ref{sec:nus}, give us a second take on the flavor problem. Much is to be gained if we can make sense of all the flavor issues---including quark mixing and neutrino mixing---together.

It is striking that, of all the parameters of the standard model (there are no fewer than  twenty-six, as listed in Table~1), at least twenty pertain to flavor, and we have no idea what determines them, nor at what energy scale they are  set\footnote{For the influence of standard-model parameters on everyday experience, see~\cite{Cahn:1996ag}.}.
In contrast, we can see how the low-energy values of the coupling parameters $\alphas$, $\alpha_{\mathrm{em}}$, $\sin^2\theta_{\mathrm{W}}$ might be set by evolution from a unified theory at a high scale.   If we succeed in establishing that the Higgs mechanism, as embodied in the electroweak theory, explains \emph{how} the masses and mixing angles arise, we still will not know \emph{why} they have the values we observe. We do not know, for example, what makes an electron an electron and a top quark a top quark. 
That is \emph{physics beyond the standard model,} even in the case of the electron mass! 
\begin{table}[t!]
\centering
{Table 1. Parameters of the Standard Model  \strut}
\begin{ruledtabular}
{\begin{tabular}{@{}cl@{}}
3 & Coupling parameters, $\alphas$, $\alpha_{\mathrm{em}}$, $\sin^2\theta_{\mathrm{W}}$ \\[3pt]
2 & Parameters of the Higgs potential \\
1 & Vacuum phase (QCD) \\[3pt]
6 & Quark masses \\
3 & Quark mixing angles \\
1 & $\mathcal{CP}$-violating quark phase\\
3 & Charged-lepton masses \\
3 & Neutrino masses \\
3 & Leptonic mixing angles \\
1 & Leptonic $\mathcal{CP}$-violating phase (+ Majorana phases?)\\
\hline
$26^+$ & Arbitrary parameters\\
\end{tabular}}
\end{ruledtabular}
\label{tab:tbl1}
\end{table}

Flavor physics is rich in unknowns:
\begin{enumerate}
\setcounter{enumi}{\value{runningcount}}
\item Have we found the ``periodic table'' of elementary particles? Is the cartoon in Figure~\ref{fig:SMcartoon} complete\footnote{Bear in mind that when Dmitri Mendeleev constructed his table of the chemical elements, the seven noble gases that make up column 18 of the modern table were unknown.}?
\item Can we find evidence of right-handed charged-current interactions? Is Nature built on a fundamentally asymmetrical plan, or are the right-handed weak interactions simply too feeble for us to have observed until now, reflecting an underlying symmetry hidden by spontaneous symmetry breaking? 
\item What is the relationship of left-handed and right-handed fermions?
\item Are there additional electroweak gauge bosons, beyond $W^\pm$ and $Z$?
\item Is charged-current universality exact? What about charged-lepton flavor universality?
\item Where are flavor-changing neutral currents? In the standard model, these are absent at tree level and highly suppressed by the Glashow--Iliopouolos--Maiani mechanism. They arise generically in proposals for physics beyond the standard model, and need to be controlled. And yet we have made no sightings\footnote{For example, the rates observed for the rare decays $K^+ \to \pi^+ \nu \bar{\nu}$ and $B_s \to \mu^+\mu^-$ are consistent with standard-model predictions.}! Why not?
{\item Can we find evidence for charged-lepton flavor violation?}
\item How well can we test the standard-model correlation among the branching fractions $\mathcal{B}(K^+ \to \pi^+\nu\bar{\nu}$) and $\mathcal{B}(B_s \to \mu^+\mu^-)$, and the quark-mixing matrix parameter $\gamma$?
\item What do quark--lepton generations mean? Is there a family symmetry?
\item Why are there three families of quarks and leptons? 
\item Are there new species of quarks and leptons, perhaps bearing exotic electric or color charges? \label{item:stuff}
\item Is there any link to a dark sector?
\item What will resolve the discrepant values of $\abs{V_{ub}}$ and $\abs{V_{cb}}$ measured in inclusive and exclusive decays?
\item Is the $3\times 3$ (Cabibbo--Kobayashi--Maskawa) quark-mixing matrix unitary?
\item Will flavor observables allow us to establish and diagnose a break in the standard model? 
\item How would a high-luminosity $e^+e^-$ collider running at the $Z^0$ pole enhance the study of heavy flavors?
\item {Do flavor parameters \emph{mean} anything at all?} 
 If flavor parameters have meaning (beyond engineering information), {what is the meta-question\footnote{According to the string-landscape point of view, the values might not have any deep significance. See, for example, \cite{Susskind:2003kw}.}?}

\setcounter{runningcount}{\value{enumi}}
\end{enumerate}
As the most massive constituent (by far!) the top quark is an object of special fascination.
Its properties and interactions touch many topics in particle physics.
\begin{enumerate}
\setcounter{enumi}{\value{runningcount}}
\item How much can we tighten the $\m_t$-$M_W$-$M_H$ constraints that test our understanding of electroweak quantum corrections?
\item Does top's large $Ht\bar{t}$ (Yukawa) coupling imply a special role in electroweak symmetry breaking? How does it influence $t\bar{t}$ dynamics?
Does its large  mass make top an outlier or, with a Yukawa coupling close to unity, the only normal fermion?
\item How well can we constrain the quark-mixing matrix element $V_{tb}$ in single-top production, or in other observables?
\item How complete is our understanding of $t\bar{t}$ production in QCD: total and differential cross sections, charge asymmetry, spin correlations, etc.? Which observables express entanglement?
\item How well can we constrain the top-quark lifetime? How {free} is the top quark?
\item Are there $t\bar{t}$ resonances? {Might $t$ hadronize a bit in the environment of heavy-ion collisions?}
\item Can we find evidence of flavor-changing top decays $t \to (Z, \gamma)(c,u)$? 
\setcounter{runningcount}{\value{enumi}}
\end{enumerate}

\section{Flavor II: Neutrino Physics \label{sec:nus}}
The discovery\footnote{For narratives of the decisive observations involving atmospheric and solar neutrinos, see \cite{Kajita:2016cak, McDonald:2016ixn}.} that neutrinos oscillate among the three known species, $\nu_e, \nu_\mu, \nu_\tau$---made, incidentally, with neutrinos from natural sources, not accelerators---is one of the great achievements of particle physics in recent decades. Several anomalies remain to be understood, as well. Accelerator-based experiments are playing an essential role in following up the discovery, and neutrino superbeams generated by meson decay at J-PARC and Fermilab will raise proton power to a megawatt or more. The mammoth DUNE and Hyper-Kamiokande detectors as well as new short-baseline experiments are expected to begin operation over the next decade\footnote{Nonaccelerator experiments that rely on reactors (such as JUNO) or natural sources (such as the neutrino telescopes IceCube and KM3Net plus Hyper-K and DUNE as observatories) enrich the neutrino campaign.}. 

A Neutrino Factory based on a muon storage ring could provide a  second act for the coming generation of accelerator-based neutrino experiments. Beyond its application to oscillation experiments~\cite{Denton:2024glz} as an intense source with known composition, an instrument that delivered $10^{20}~\nu$ per year could be a highly valuable resource for on-campus experiments. Neutrino interactions on thin targets, polarized targets, or active targets could complement the nucleon-structure programs carried out in electron scattering at Jefferson Lab and elsewhere. \cite{Ball:2000qd,Bogacz:2022xsj,deGouvea:2025zfq} Is such a  prospect truly interesting, or merely amazing?

Among the questions we would like to answer are these: 
\begin{enumerate}
\setcounter{enumi}{\value{runningcount}}
\item In what way is neutrino mass a sign of physics beyond the standard model?
\item Do three light (left-handed) neutrinos suffice? How well can we test that the $3\times 3$ Pontecorvo--Mak--Nakagawa--Sakata (PMNS) neutrino-mixing matrix is unitary?
\item What is the order of levels of the mass eigenstates $\nu_1, \nu_2, \nu_3$? It is known that the $\nu_e$-rich $\nu_1$ is the lighter of the ``solar pair,'' with the more massive $\nu_2$. Does the $\nu_e$-poor $\nu_3$ lie above (``normal'' mass hierarchy) or below (``inverted'' hierarchy) the others?
\item What is the absolute  scale of neutrino masses? What might we learn if  a conflict arose between cosmological inferences and kinematic determinations from $^3\mathrm{H}$ $\beta$-decay (KATRIN) or $^{163}\mathrm{Ho}\to\!{^{163}\mathrm{Dy}}$ electron capture?
\item Does the see-saw hypothesis explain the smallness of neutrino masses?
\item What is the flavor composition of $\nu_3$? Is it richer in $\nu_\mu$ or $\nu_\tau$\footnote{Current neutrino oscillation data suggest that $\theta_{23}$ is not exactly 45 degrees, but rather lies either slightly below or slightly above this value. This ambiguity is known as the octant degeneracy.}?
\item Is $\mathcal{CP}$ symmetry violated in neutrino oscillations? To what degree?
\item Will neutrinos give us insight into the matter excess in the universe (through leptogenesis)?
\item Are neutrinos Majorana particles? While this issue is primarily addressed by searches for neutrinoless double-$\beta$ decay, collider searches for same-sign lepton pairs also speak to it.
\item Are there light sterile neutrinos that lack \ewgg\ charges? If so, how could they arise?
\item Do neutrinos have nonstandard interactions, beyond those mediated by $W^\pm$ and $Z$? If $\nu_e$ were found to have novel interactions, how could we follow up?
\item What might we learn from neutrino observatories about astrophysical objects and processes?
\item Are all the neutrinos stable?
\item How can we detect the cosmic neutrino background? If stable, the three neutrino species  and their corresponding antineutrinos should each be present in the current universe with a number density of $\approx 56\cm^{-3}$ and a temperature of $T_\nu \approx 2\K \approx 1.7\times 10^{-4}\ev$. 
\item Do neutrinos contribute appreciably to the dark matter component of the universe?
\item What physics opportunities are presented by TeV neutrinos from LHC interaction points or a future muon collider? How might they surpass the $e^\mp p$ studies from HERA at $\sqrt{s} \approx 320\gev$,  equivalent to 50-TeV electrons incident on stationary protons?
\item What opportunities to discover physics beyond the standard model will the next generation of long-baseline neutrino experiments open~\cite{DUNE:2020fgq}?
\setcounter{runningcount}{\value{enumi}}
\end{enumerate}

\section{Don't Forget the Strong Interactions!\label{sec:strong}}
Quantum Chromodynamics is the basis of hadronic physics. The fundamental fields of the theory, quarks and gluons, are manifest in proton structure, which is to say in high-resolution, hard-scattering studies, in the exploration of matter at high density, and in lattice calculations. The effective degrees of freedom of the theory show themselves as constituent quarks and Goldstone bosons and in effective field theories, isobar (resonance) models, and nuclei and nuclear structure.

We aspire to {compute} the properties of hadrons, explain the absence of unseen species, and predict the existence of new varieties of hadrons; explain why quarks and gluons are not observed as free particles; derive the interactions among hadrons as a collective effect of the interactions among constituents.

Collider experiments have stimulated heroic progress in perturbative methods to elaborate the consequences of quantum chromodynamics for scattering experiments. Lattice QCD has broken new ground in understanding the strong-coupling regime.  It may provide insights that complement those gleaned from experiment, by giving precise form to our intuitive pictures of hadrons. Modulo the strong $\mathcal{CP}$ problem, QCD could be structurally complete and consistent up to the Planck energy, $1.2 \times 10^{19}\gev$---but that doesn't prove it must be. We must prepare for surprises, including  quark compositeness (Question~\ref{item:comp}) or new kinds of colored matter (Question~\ref{item:stuff}), as well as deviations from our understanding of QCD itself. 

In some contexts, it may seem that the strong interactions are chiefly of interest as backgrounds to searches for ``new physics.'' That view is far too narrow; among the issues we would like to resolve are these:
\begin{enumerate}
\setcounter{enumi}{\value{runningcount}}
\item Why is isospin a good symmetry? What does it mean? Is it accidental?
\item Where is colorspin $\mathrm{SU(6)_{cs}\equiv SU(3)_c \otimes SU(2)_{spin}}$ a useful organizing principle?
\item To what degree might we derive the parton model from first-principles quantum chromodynamics for lepton--hadron and hadron--hadron collisions? In what circumstances does factorization hold?
\item What are the limitations of  the one-dimensional $\infty$-momentum-frame parton model?
\item What can generalized parton distributions (GPDs) and Transverse Momentum Distributions (TMDs) teach us about the static properties and interactions of nucleons and other hadrons?
\item Will high-energy collisions reveal new phenomena within QCD?  Can we use machine learning to characterize untriggered or pile-up events beyond the traditional classes of diffraction and short-range order\footnote{Exploring the dependence on total and  heavy-flavor multiplicity and comparing $pp$, $p$A, and AA collisions should be revealing. I explored some possibilities in~\cite{Quigg:2010nn}.}?
\item How will correlations among partons in a proton manifest themselves?
\item Under what circumstances is the notion of a compact diquark informative? How fruitful is the analogy between heavy--light $(Q\bar{q})$ mesons and doubly heavy $(QQq)$ baryons?
\item What role do diquarks play in color--flavor locking, color superconductivity, etc.?
\item Might event structures distinguish spatial configurations of partons within protons, such as three separated quarks, quark--diquark, compact triquark, or flux tube? Are there any useful analogies to nucleus--nucleus collisions?
\item How are charge, flavor, etc., distributed within the proton,  neutron, and pseudoscalar mesons?
\item Where does nucleon spin reside?
\item How can concepts developed for heavy-ion collisions inform our understanding of $pp$ collisions under extreme conditions?
\item Will any surprises arise in ``dead-cone'' studies using boosted heavy quarks ($t, b, c$)?
\item How will the high density of ``wee partons'' that carry negligible momentum fractions affect $pp$ collisions? How will gluon saturation manifest itself? What will be other consequences?
\item What is the importance of intrinsic heavy flavors? Might the concept extend beyond quarks?
\item Can we develop a comprehensive understanding of hadron structure, including body plans beyond $qqq$ and $q\bar{q}$?
\item How do $6q$ body plans prefigure the emergence of nuclear structure?
\item Do nuclear bound states exist in the chiral limit? What range of quark masses might lead to periodic tables resembling the one we know?
\item What lessons for hadron spectroscopy can we draw from applications of supersymmetry to nuclei?
\item Do observed patterns among $(q\bar{q})$ mesons, $(qqq)$ baryons, tetraquarks, and pentaquarks reflect heavy-quark symmetry mass formulas such as  
$M(Q_iQ_j\bar{q}_k\bar{q}_\ell)\!-\!M(Q_iQ_jq_m)$\\
$\phantom{MMMMM}\qquad\qquad =M(Q_xq_kq_\ell)\!-\!M(Q_x\bar{q}_m)$?
\item What is the nature of the Pomeron? What can we learn from events with large rapidity gaps?
\item How are the confinement scale, the scale of chiral symmetry breaking, and the strong coupling parameter $\Lambda_\mathrm{QCD}$ related?
\item Do (nearly) pure glueball states exist?
\item What new phenomena occur in the strong interaction at high fermion density? How far can we extend the QCD phase diagram? What new insights might that bring to astrophysical phenomena such as neutron stars?
\item Can we \emph{{prove}} that quantum chromodynamics confines color\footnote{A proof of confinement, in the form of a mass gap in Yang--Mills theory, is one of seven Millennium Problems set  by the Clay Mathematics Institute~\cite{Massgap}.}? Will free quarks (or other colored objects) be observed?
\item What resolves the strong $\mathcal{CP}$ problem? 
\item How can we search comprehensively for axions and axion-like particles?
\setcounter{runningcount}{\value{enumi}}
\end{enumerate}

\section{Unified Theories \label{sec:GUTS}}
The standard model based on \smgg\ gauge symmetry encapsulates much 
of what we know and describes many observations, but it leaves many 
things unexplained. Both the success and the incompleteness of the 
standard model encourage us to look beyond it to a more comprehensive 
understanding. One attractive way to proceed is by \textit{enlarging 
the gauge group,} which we may attempt either by accreting  new 
symmetries or by unifying the symmetries we have already recognized.

Left-right symmetric  models,  based on  gauge 
symmetries such as $\mathrm{SU(3)_{c} \otimes SU(2)_{L}\otimes SU(2)_{R} \otimes 
U(1)}_{B-L},$ where $B$ and $L$ stand for Baryon and Lepton numbers, follow the first path. Such models attribute the 
observed parity violation in the weak interactions to spontaneous 
symmetry breaking---the $\mathrm{SU(2)_{R}}$ (right-handed weak isospin) symmetry is broken at a 
higher scale than the $\mathrm{SU(2)_{L}}$---and naturally 
accommodate Majorana neutrinos. Left-right symmetric 
theories also open new possibilities, including transitions that 
induce $n \leftrightarrow \bar{n}$ oscillations and a mechanism for 
spontaneous $\mathcal{CP}$ violation. More generally, enlarging the 
gauge group by accretion seeks to add a missing element or to explain 
additional observations.

Unified theories, on the other hand, seek to find a symmetry group 
${\mathcal{G}} \supset \smgg$
(usually a simple group, to maximize the predictive power) that 
contains the known interactions. This approach is motivated by the 
desire to unify quarks and leptons and to reduce the number of 
independent coupling constants, the better to understand the relative 
strengths of the strong, weak, and electromagnetic interactions at 
laboratory energies. Supersymmetric unified theories bring the added ambitions of 
incorporating gravity and joining constituents and forces.

Two very potent ideas are at play here. The first is the idea of 
unification itself: what Feynman calls \textit{amalgamation,} which is 
the central notion of \textit{generalization and synthesis} that scientific 
explanation represents. Examples from the history of physics include 
Maxwell's joining of electricity and magnetism and light; the atomic 
hypothesis, which places thermodynamics and statistical mechanics 
within the realm of Newtonian mechanics; and the links among atomic 
structure, chemistry, and quantum mechanics.
The second is the notion---further developed in \S\ref{sec:newscale}---that the human scale of space and time is not
privileged for understanding Nature, and may even be disadvantaged.

These considerations lead us toward a more complete electroweak 
unification, which is to say a simple $\mathcal{G} \supset \ewgg$; a 
quark-lepton connection; a ``grand'' unification of the strong, weak, 
and electromagnetic interactions, based on a simple group 
$\mathcal{G} \supset \smgg$. If we choose the task of grand 
unification, we must find a group that contains the known interactions 
and that can accommodate the known fermions---either as one generation 
plus replicas, or as all three generations at once. The unifying group 
will surely contain interactions beyond the established ones, and we 
should be open to the possibility that the fermion representations 
require the existence of particles yet undiscovered.

\begin{enumerate}
\setcounter{enumi}{\value{runningcount}}
\item Quarks and leptons are structureless spin-$\half$ particles  that come in matched sets, 
as required by anomaly cancellation for a renormalizable \ewgg\ theory.
 What is the relationship of quarks to leptons?
\item Is it productive to regard lepton number as the ``fourth color?'' 
\item Why is charge quantized?  [$Q_{d}=\frac{1}{3}Q_{e}$,
$Q_{p}+Q_{e}=0$, $Q_{\nu}-Q_{e}=Q_{u}-Q_{d}$,
$Q_{\nu}+Q_{e}+3Q_{u}+3Q_{d}=0$.]
\item What is the meaning of electroweak universality, embodied in the
matching left-handed doublets of quarks and leptons?
\item Are there new gauge interactions that link quarks with leptons? If so, which quark doublet (or mixture) is matched with which lepton doublet?
Are there other sources of lepton- and baryon-number violation?
\item What is the (grand) unifying symmetry?
\item What determines the low-energy gauge symmetries?
\item What are the steps to unification? One more, or multiple?
\item Can the three distinct coupling parameters of the standard model
($\alphas, \alpha_{\mathrm{em}}, \sin^{2}\theta_{W}$) or ($g_{\mathrm{s}}, g,
g^{\prime}$) be reduced to two or one\footnote{Here $g$ and $g^\prime$ are respectively the couplings of the $\mathrm{SU(2)_L}$ and $\mathrm{U(1)}_Y$ interactions.}? Is perturbation theory a reliable guide to coupling unification?
Is there a unification point where all (suitably defined) couplings
coincide?
\item How can we define a nonperturbative ultraviolet regulator for chiral gauge theories such as the standard model?
\item What sets the mass scale for the additional gauge bosons in a unified theory? \ldots for the additional Higgs bosons?
\item Is the proton unstable? If so, how does it decay?
\item Is neutron--antineutron oscillation observable?
\item Are there millicharged particles?  Other signs of additional U(1) gauge symmetries?
\item How can we incorporate gravity?
\item Why is gravity so weak?
\item To what distance scale (small or large) does the inverse-square law of gravitation hold? 
\item What is the nature of spacetime?  Is it emergent?  How many dimensions?
\setcounter{runningcount}{\value{enumi}}
\end{enumerate}

\section{Where Is the Next Important Scale? \label{sec:newscale}}
A popular---and productive---way to introduce the grand sweep of science is to show what the world looks like when viewed at various magnifications~\cite{Eames:POT}. Colleagues who spend time at CERN may be familiar with the 200 Swiss Franc note\footnote{The \href{https://www.snb.ch/en/the-snb/mandates-goals/cash/new-banknotes/design/200-franc-note}{CHF200 note}, issued 22 August 2018, illustrates the theme of scientific expertise, depicting the right-hand rule and a particle collision. }.
On close inspection, the security strip reveals an abstract map of the geological ages of Switzerland with a timeline\footnote{The timeline was constructed with the advice of our colleague \href{https://inspirehep.net/authors/1011717?ui-citation-summary=true}{G\"unther Dissertori}.} of milestones in the evolution of the universe, from the Planck epoch, ($t_{\text{Pl}}  \approx  10^{-43}\s)$\footnote{Max Planck defined ``natural units of measurement'' in terms of a few fundamental constants~\cite{*[{}][{. The numerical values he quotes are $2\pi\times$ the modern values because his definitions used $h$ rather than $\hbar$. For an insightful commentary, see~\cite{10.1063/1.2138392}}]PlanckUnits:1899}: 
\begin{align}
\text{Length:}\quad L_{\text{Pl}} & \equiv  \sqrt{\hbar G_{\text{N}}/c^3} \approx 1.6 \times 10^{-35}\m \nonumber \\
\text{Mass:}\quad M_{\text{Pl}} & \equiv \sqrt{\hbar c/ G_{\text{N}}} \approx 1.2 \times 10^{19}\gevcc \nonumber \\ 
\text{Time:}\quad t_{\text{Pl}} & \equiv \sqrt{\hbar G_{\text{N}}/c^5} \approx 5.4 \times 10^{-44}\s\nonumber
\end{align}
 Here $G_{\text{N}} = 6.708 83(15)\times10^{-39} \,\hbar c\, (\text{GeV}/c^2)^{-2}$ is Newton's gravitational constant. A little perspective---the gigantic Planck mass is only about 22 micrograms in macroscopic units.}, to the current age of the Universe, $13.8\gy \approx 4.35 \times 10^{17}\s$~\cite{Planck:2018vyg}.

The timeline's sixty orders of magnitude do not span the full range of our observations---or our imagination. Perhaps Planck time is a lower limit; I for one find it hard to think rationally about the universe at earlier times. But at the upper end, our observations extend well beyond the current age of the universe. The measured rates\footnote{Half-life $T_{1/2}$ is related to the mean life conventional in particle physics by $\tau = T_{1/2}/\ln{2}$.} for double-$\beta$ decay and double electron capture,
\begin{align}
T_{1/2}(^{136}\text{Xe}_{\beta\beta\nu\nu}) & \approx 2.2 \times 10^{21}\y~\text{\cite{EXO-200:2013xfn}}\nonumber \\
T_{1/2}(^{124}\text{Xe}_{\text{ECEC}\nu\nu}) & \approx 1.1 \times 10^{22}\y~\text{\cite{XENON:2022evz}}\nonumber
\end{align}
add a dozen more orders of magnitude.  Lower bounds on the electron and proton lifetimes add at least six more. We can contemplate corresponding spans in energy and distance.

The point of this discussion is not merely to be dazzled by large numbers. Within physics, the utility of multiple scales has grown in significance since the quantum-mechanical revolution of the 1920s.  To
understand why a rock is solid, or why a metal gleams, we must
discern its structure on a scale a billion  times smaller 
than the human scale, and we must learn the rules that prevail
there\footnote{A moment's reflection will reveal parallels throughout science. 
Studies in biology, cosmology, and ecology  range over distances and time spans both exponentially larger and exponentially smaller than the human scale.}.

Other scales may well be privileged for
understanding certain globally important aspects of the universe. For example, a unification scale might be privileged for understanding how it came to be that the fine structure constant $\alpha \approx 1/137$ in the long-wavelength limit and the strong coupling constant $\alphas \approx 1/5$ at energies characteristic of the $\Upsilon$ resonances;  or why fermion masses exhibit the (seemingly irrational) pattern they do. What has matured recently is our comfort at cruising between different scales of momentum or distance and our understanding---through the renormalization group and effective field theories---of how one scale relates to another. 

I believe that the discovery that \emph{the human scale is not 
preferred} is as important as the discoveries that 
the human location is not privileged (Copernicus) and that there is no 
preferred inertial frame (Einstein), and will prove to be as influential. I consider this insight to be The Great Lesson of Twentieth-Century Science.

In the recent past, we had the good fortune to identify the \onetev\ as the place to find new physics, specifically to elucidate the nature of electroweak symmetry breaking. We do not (yet!) have a similar target for the next step. 

The puzzle of the next important scale is wide-open. Here are a few questions to encourage further thought:
\begin{enumerate}
\setcounter{enumi}{\value{runningcount}}
\item At what scale are the Yukawa couplings that determine (we think!) charged-fermion masses set? What about neutrino masses?
\item Can we establish experimental evidence for  a unification scale, perhaps  $\sim 10^{15\,\text{-}16}\gev$, at which the strong, weak and eletromagnetic interactions are unified? Might there be an intermediate scale for another partial unification?
\item Will new physics appear at $1 \times, 10 \times, 100 \times, \ldots$ the electroweak scale? (I hope we don't have to wait until the Planck scale!)
\setcounter{runningcount}{\value{enumi}}
\end{enumerate}

\section{The Cosmic Connection \label{sec:Astro/Cosmo}}
A century ago, Cecilia Payne deduced the relative abundance of the elements by studying the spectra of stellar atmospheres~\cite{Payne:1925}.
By mid-century, as memorably recounted by Steven Weinberg~\cite{Weinberg:1977ji}, astrophysicists had established the essential features of primordial nucleosynthesis. Today, we consider the universe both as a testing ground for our understanding of the laws of Nature, and as a source of challenges to the completeness of our worldview.  

In the years since the discovery of the cosmic microwave background~\cite{Penzias:1965wn,Dicke:1965zz}, cosmologists have developed, revised, and refined a standard cosmological model that is impressive in its simplicity and scope~\cite{*[{For a recent compact summary, see }][]OlivePeacock}. The {young} universe was very hot and dense and has cooled as it expanded to its current state. At very {early} times, around $10^{-35}\s$, cosmologists surmise that the universe experienced a period of rapid, roughly exponential, growth called the inflationary epoch. 

Within the framework of General Relativity, the post-inflation expansion rate is controlled by the mass-energy content of the universe. At the current epoch, indicated by the subscript $0$, the principal components are matter $\Omega_{\text{m}0} \approx 0.315$, of which ordinary baryonic matter makes up $\Omega_{\text{b}0} \approx 0.05$ and (unidentified) dark matter $\Omega_{\text{DM}0} \approx 0.265$, and a ``dark energy,'' $\Omega_{\text{DE}0} \approx 0.685$, that impels the observed accelerating expansion of the universe~\cite{SupernovaSearchTeam:1998fmf,SupernovaCosmologyProject:1998vns}\footnote{For an early review of the implications of accelerating expansion, see~\cite{Frieman:2008sn}.}. These energy densities are measured with respect to the critical density for which a flat, matter-dominated, expanding universe would cease its expansion asymptotically after an infinite time, $\rho_{\text{c}} \equiv  3H_0^2 / 8\pi G_{\text{N}}$. Here $H_0$ is the Hubble expansion parameter measured in the current universe as $\approx 72
\text{ km\s}^{-1} \text{Mpc}^{-1}$.\footnote{The Hubble parameter is defined as the logarithmic time derivative of the cosmological scale factor: $H \equiv \dot{R}/R$.} The two copious relics of the early universe, photons at 2.73\K\ and light neutrinos at 1.95\K, contribute negligibly to the current energy density.\footnote{Observations indicate that the universe is flat, i.e., without intrinsic curvature. See~\cite{Lahav:2024npe} for the current status of the cosmological parameters.} 

It is convenient to define the dimensionless scale factor, $a = R/R_{0}$, and to express the normalized energy densities of matter, radiation, and dark/vacuum 
energy as functions of $a$. The normalized densities of matter and radiation scale as
$\rho_{m}/\rho_{c} = \Omega_{m0}/a^{3}$, and $\rho_{\gamma}/\rho_{c} = \Omega_{\gamma0}/a^{4}$.
Provisionally interpreting  the dark energy component  in terms of a cosmological constant, $\Lambda$,  we have $\rho_{\Lambda}/\rho_{c} = \Omega_{m0}$, a time-independent energy density. In this picture, the universe has been successively dominated by radiation, matter, and dark energy. When radiation and matter prevail, the expansion of the universe slows; when dark energy (here represented by the cosmological constant) prevails, the expansion accelerates.

Seeking an aphorism {to express how Newton's third law links gravitation and  spacetime,} John Wheeler wrote~\cite{*[{}][{. See p.~235.}]Wheeler:1998},
``Spacetime tells matter how to move; matter tells spacetime how to curve.'' 
Following Wheeler's lead, we might say that ``Expansion tells the matter and energy content how to evolve, content tells the expansion how to proceed.''

After centuries of observation and study of the universe, we have come to this tantalizingly successful but incomplete standard cosmology. There is much that we do not know, including the complete thermal history (to now) of the universe and the detailed composition of the universe at large. To progress, we must learn to read new strata of the cosmological fossil record, refine the precision of our observations, and develop new ways of looking at the universe. The detection of gravitational radiation enriches multimessenger astronomy and is a powerful addition to the cosmological toolkit. Some timely questions that may lead to further refinements of our cosmological narrative include: 
\begin{enumerate}
\setcounter{enumi}{\value{runningcount}}
\item To what degree does the cosmological principle---the notion that the universe is homogeneous and isotropic on the largest scales---hold?
\item How can we refine our knowledge of primordial nuclear abundances, with their implications for the constituents of the early universe?
\item How precisely does  the cosmic microwave background radiation follow a blackbody distribution?
\item What accounts for the predominance of matter over antimatter in the universe?
\item Is there a single species of dark matter, or many?
\item Are there any alternatives/complements to collisionless dark matter? Does our understanding of gravitation require revision?
\item Does the dark energy density evolve with the cosmic scale factor? Is the dark energy density properly characterized by a cosmological constant $\Lambda$ or by a dynamical mechanism? If a cosmological constant, what sets the scale? If the dark energy density is not constant ($\propto a^{-3(1+w)}$), what is the dark energy equation of state, $w$? 
\item Is there a dynamical interplay between cosmological evolution and scalar-field  dynamics (perhaps including the Higgs field)?
\item Can we establish the existence of an inflationary epoch? If so, what triggered it?
\item Is the Hubble tension ($H_0^{\mathrm{local}} - H_0^{\mathrm{Planck}}) \approx (4 \hbox{--} 6)\sigma$ real? If so, what does it teach us?
\item Can we probe dark energy in laboratory experiments?
\item How can technologies developed for accelerators advance the search for feebly interacting particles? 
What might be the role of quantum technologies?
\setcounter{runningcount}{\value{enumi}}
\end{enumerate}

%
%
%
%
%
%
%
%
%
%

\section{Final Remarks}
The progress of accelerator science and technology has driven the development of particle physics, while the imperatives of experimental research have stimulated advances in accelerator research. I am confident that the synergy will continue. The  new machines under discussion---circular $e^+e^-$ Higgs factory, $e^+e^-$ linear collider, ``100-TeV'' hadron collider, electron--hadron collider, muon collider, etc.---have compelling scientific motivations and appear achievable: they will require significant technological progress, but they do not demand cascading miracles\footnote{These machines do lie near the edge of practicality in terms of performance and resources required \cite{Bloise:2025buj,Bloom:2022gux,Bloom:2024ixe}. It is liberating---and important---for us to look beyond projects we can credibly propose today and to dream for the far future. A good model can be found in James Bjorken's 1982 lectures on storage rings to attain $\sqrt{s}=1000\tev$~\cite{Bjorken:1982iw}! }. 

Together with the detectors that our experimental colleagues mount to exploit them, they are exemplars of the most amazing achievements of human beings---all the more admirable for being dedicated to the advancement of knowledge. We do not seek to build these machines out of mere habit, but because the scientific frontiers they will open are incredibly exciting. While we do our best to predict what new understanding the next accelerators will yield, the real thrill is that we don't know what we will find\footnote{As a recent short subject puts it, ``Real researchers don't know what they're looking for.'' \cite{VraisChercheurs}}.

I suspect that every generation has wondered whether the next machine will be the last. Even if timely innovations in the past have pushed the boundaries of what we can do, that anxiety will always be present. The size, complexity, cost, and time scale of the accelerators we would like to attempt next amplify the concern. We will not execute all of these ideas. They will compete for resources, and for our enthusiasm. But it is better to have too many appealing ideas than too few!

The long duration of projects that may not come to fruition means that the particle-physics community has a special responsibility to nurture the careers of accelerator designers and builders. That responsibility falls naturally to the great laboratories, but more university physics and engineering departments should see accelerator science as a fertile intellectual discipline, with lively connections to many other fields~\cite{AFAF}. Breakthroughs and refinements in accelerator technology may find their first---or most consequential---applications far from the frontiers that preoccupy particle physicists.

Now, let us look over the horizon:

\begin{enumerate}
\setcounter{enumi}{\value{runningcount}}
\item Imagine the possibilities if wake-field acceleration or some other innovation would allow us to reach gradients of many GeV---even a TeV---per meter. How would we first apply that bit of magic, and what characteristics other than gradient would be required? 
\item If we \emph{could} shrink the dimensions of multi-TeV accelerators, is there any prospect for shrinking the dimensions of detectors that depend on particle interactions with matter? 
\item What could we do with a low-emittance, high-intensity muon source? 
\item What inventions would it take to accelerate beams of particles with picosecond lifetimes? 
\item How can we imagine going far beyond today's capabilities for steering beams? 
\item How could we apply high-transmissivity crystal channeling, if we could perfect it? 
\item How would optimizations change if we  were able to shape superconducting magnet coils out of biplanar graphene or an analogous material? 
\setcounter{runningcount}{\value{enumi}}
\end{enumerate}

In the Prologue (\S\ref{sec:invite}) to this essay, I expressed the wish that the open questions I pose would stimulate each of us to  think anew about how we---as individual researchers and as a worldwide community---can best advance our science. A healthy intellectual ecosystem in which researchers at all stages of their careers may thrive will be built of projects large and small, of long duration and short,  addressing questions great and small. We should at all times be prepared to re\"examine our priorities. I believe that this is sound strategy for individuals as well as communities.

As you make your own assessments, I encourage you to keep in mind these meta-questions: What deep questions have been with us for so long that they are less prominent in ``top-ten'' lists {(or even top-\arabic{runningcount}-and-counting lists)} than they deserve to be?
 How are we prisoners of conventional thinking, and how can we break out?
 And the most important question of all: What do we know that is not true?
\section*{Acknowledgments}
For thoughtful comments, questions, and suggestions, I am grateful to seminar participants at  Fermilab, University of Chicago, Yale University, \'{E}cole Normale Sup\'{e}rieure (Paris), University of Michigan, Jefferson Lab, Indiana University, University of Oxford, The Ohio State University, NIKHEF (Amsterdam), Universit\"{a}t Heidelberg, University of Minnesota, University of Illinois, and CERN.  Their engagement helped me clarify my intent and  shape the presentation. 

For their perceptive observations on the manuscript, I thank Gabriela Barenboim, Innes Bigaran, Anne-Katherine Burns, Joel Butler, Bogdan Dobrescu, Stefan Hoeche, Hank Lamm, Pedro Machado, and Mike Wagman. {I welcome additional advice!}

Fermilab is operated by Fermi Forward Discovery Group, LLC,
under Contract No. 89243024CSC000002 with the U.S. Department of Energy,
Office of Science, Office of High Energy Physics.

\bibliography{PQ2025}

\begin{thebibliography}{68}%
\makeatletter
\providecommand \@ifxundefined [1]{%
 \@ifx{#1\undefined}
}%
\providecommand \@ifnum [1]{%
 \ifnum #1\expandafter \@firstoftwo
 \else \expandafter \@secondoftwo
 \fi
}%
\providecommand \@ifx [1]{%
 \ifx #1\expandafter \@firstoftwo
 \else \expandafter \@secondoftwo
 \fi
}%
\providecommand \natexlab [1]{#1}%
\providecommand \enquote  [1]{``#1''}%
\providecommand \bibnamefont  [1]{#1}%
\providecommand \bibfnamefont [1]{#1}%
\providecommand \citenamefont [1]{#1}%
\providecommand \href@noop [0]{\@secondoftwo}%
\providecommand \href [0]{\begingroup \@sanitize@url \@href}%
\providecommand \@href[1]{\@@startlink{#1}\@@href}%
\providecommand \@@href[1]{\endgroup#1\@@endlink}%
\providecommand \@sanitize@url [0]{\catcode `\\12\catcode `\$12\catcode
  `\&12\catcode `\#12\catcode `\^12\catcode `\_12\catcode `\%12\relax}%
\providecommand \@@startlink[1]{}%
\providecommand \@@endlink[0]{}%
\providecommand \url  [0]{\begingroup\@sanitize@url \@url }%
\providecommand \@url [1]{\endgroup\@href {#1}{\urlprefix }}%
\providecommand \urlprefix  [0]{URL }%
\providecommand \Eprint [0]{\href }%
\providecommand \doibase [0]{http://dx.doi.org/}%
\providecommand \selectlanguage [0]{\@gobble}%
\providecommand \bibinfo  [0]{\@secondoftwo}%
\providecommand \bibfield  [0]{\@secondoftwo}%
\providecommand \translation [1]{[#1]}%
\providecommand \BibitemOpen [0]{}%
\providecommand \bibitemStop [0]{}%
\providecommand \bibitemNoStop [0]{.\EOS\space}%
\providecommand \EOS [0]{\spacefactor3000\relax}%
\providecommand \BibitemShut  [1]{\csname bibitem#1\endcsname}%
\let\auto@bib@innerbib\@empty
\bibitem [{\citenamefont {Aad}\ \emph {et~al.}(2012)\citenamefont {Aad} \emph
  {et~al.}}]{ATLAS:2012yve}%
  \BibitemOpen
  \bibfield  {author} {\bibinfo {author} {\bibnamefont {Aad}, \bibfnamefont
  {Georges}},  \emph {et~al.} (\bibinfo {collaboration} {ATLAS})} (\bibinfo
  {year} {2012}),\ \bibfield  {title} {\enquote {\bibinfo {title} {{Observation
  of a new particle in the search for the Standard Model Higgs boson with the
  ATLAS detector at the LHC}},}\ }\href {\doibase
  10.1016/j.physletb.2012.08.020} {\bibfield  {journal} {\bibinfo  {journal}
  {Phys. Lett. B}\ }\textbf {\bibinfo {volume} {716}},\ \bibinfo {pages}
  {1--29}},\ \Eprint {http://arxiv.org/abs/1207.7214} {arXiv:1207.7214
  [hep-ex]} \BibitemShut {NoStop}%
\bibitem [{\citenamefont {Aad}\ \emph {et~al.}(2022)\citenamefont {Aad} \emph
  {et~al.}}]{ATLAS:2022vkf}%
  \BibitemOpen
  \bibfield  {author} {\bibinfo {author} {\bibnamefont {Aad}, \bibfnamefont
  {Georges}},  \emph {et~al.} (\bibinfo {collaboration} {ATLAS})} (\bibinfo
  {year} {2022}),\ \bibfield  {title} {\enquote {\bibinfo {title} {{A detailed
  map of Higgs boson interactions by the ATLAS experiment ten years after the
  discovery}},}\ }\href {\doibase 10.1038/s41586-022-04893-w} {\bibfield
  {journal} {\bibinfo  {journal} {Nature}\ }\textbf {\bibinfo {volume}
  {607}}~(\bibinfo {number} {7917}),\ \bibinfo {pages} {52--59}},\ \bibinfo
  {note} {[Erratum: Nature 612, E24 (2022)]},\ \Eprint
  {http://arxiv.org/abs/2207.00092} {arXiv:2207.00092 [hep-ex]} \BibitemShut
  {NoStop}%
\bibitem [{\citenamefont {Aad}\ \emph {et~al.}(2024)\citenamefont {Aad} \emph
  {et~al.}}]{ATLAS:2024lyh}%
  \BibitemOpen
  \bibfield  {author} {\bibinfo {author} {\bibnamefont {Aad}, \bibfnamefont
  {Georges}},  \emph {et~al.} (\bibinfo {collaboration} {ATLAS})} (\bibinfo
  {year} {2024}),\ \bibfield  {title} {\enquote {\bibinfo {title}
  {{Interpretations of the ATLAS measurements of Higgs boson production and
  decay rates and differential cross-sections in pp collisions at $ \sqrt{s} $
  = 13 TeV}},}\ }\href {\doibase 10.1007/JHEP11(2024)097} {\bibfield  {journal}
  {\bibinfo  {journal} {JHEP}\ }\textbf {\bibinfo {volume} {11}},\ \bibinfo
  {pages} {097}},\ \Eprint {http://arxiv.org/abs/2402.05742} {arXiv:2402.05742
  [hep-ex]} \BibitemShut {NoStop}%
\bibitem [{\citenamefont {Abi}\ \emph {et~al.}(2021)\citenamefont {Abi} \emph
  {et~al.}}]{DUNE:2020fgq}%
  \BibitemOpen
  \bibfield  {author} {\bibinfo {author} {\bibnamefont {Abi}, \bibfnamefont
  {B}},  \emph {et~al.} (\bibinfo {collaboration} {DUNE})} (\bibinfo {year}
  {2021}),\ \bibfield  {title} {\enquote {\bibinfo {title} {{Prospects for
  beyond the Standard Model physics searches at the Deep Underground Neutrino
  Experiment}},}\ }\href {\doibase 10.1140/epjc/s10052-021-09007-w} {\bibfield
  {journal} {\bibinfo  {journal} {Eur. Phys. J. C}\ }\textbf {\bibinfo {volume}
  {81}}~(\bibinfo {number} {4}),\ \bibinfo {pages} {322}},\ \Eprint
  {http://arxiv.org/abs/2008.12769} {arXiv:2008.12769 [hep-ex]} \BibitemShut
  {NoStop}%
\bibitem [{\citenamefont {Acharya}\ \emph {et~al.}(2014)\citenamefont {Acharya}
  \emph {et~al.}}]{MoEDAL:2014ttp}%
  \BibitemOpen
  \bibfield  {author} {\bibinfo {author} {\bibnamefont {Acharya}, \bibfnamefont
  {B}},  \emph {et~al.} (\bibinfo {collaboration} {MoEDAL})} (\bibinfo {year}
  {2014}),\ \bibfield  {title} {\enquote {\bibinfo {title} {{The Physics
  Programme Of The MoEDAL Experiment At The LHC}},}\ }\href {\doibase
  10.1142/S0217751X14300506} {\bibfield  {journal} {\bibinfo  {journal} {Int.
  J. Mod. Phys. A}\ }\textbf {\bibinfo {volume} {29}},\ \bibinfo {pages}
  {1430050}},\ \Eprint {http://arxiv.org/abs/1405.7662} {arXiv:1405.7662
  [hep-ph]} \BibitemShut {NoStop}%
\bibitem [{\citenamefont {Aghanim}\ \emph {et~al.}(2020)\citenamefont {Aghanim}
  \emph {et~al.}}]{Planck:2018vyg}%
  \BibitemOpen
  \bibfield  {author} {\bibinfo {author} {\bibnamefont {Aghanim}, \bibfnamefont
  {N}},  \emph {et~al.} (\bibinfo {collaboration} {Planck})} (\bibinfo {year}
  {2020}),\ \bibfield  {title} {\enquote {\bibinfo {title} {{Planck 2018
  results. VI. Cosmological parameters}},}\ }\href {\doibase
  10.1051/0004-6361/201833910} {\bibfield  {journal} {\bibinfo  {journal}
  {Astron. Astrophys.}\ }\textbf {\bibinfo {volume} {641}},\ \bibinfo {pages}
  {A6}},\ \bibinfo {note} {[Erratum: Astron.Astrophys. 652, C4 (2021)]},\
  \Eprint {http://arxiv.org/abs/1807.06209} {arXiv:1807.06209 [astro-ph.CO]}
  \BibitemShut {NoStop}%
\bibitem [{\citenamefont {Albert}\ \emph {et~al.}(2014)\citenamefont {Albert}
  \emph {et~al.}}]{EXO-200:2013xfn}%
  \BibitemOpen
  \bibfield  {author} {\bibinfo {author} {\bibnamefont {Albert}, \bibfnamefont
  {J~B}},  \emph {et~al.} (\bibinfo {collaboration} {EXO-200})} (\bibinfo
  {year} {2014}),\ \bibfield  {title} {\enquote {\bibinfo {title} {{Improved
  measurement of the $2\nu\beta\beta$ half-life of $^{136}$Xe with the EXO-200
  detector}},}\ }\href {\doibase 10.1103/PhysRevC.89.015502} {\bibfield
  {journal} {\bibinfo  {journal} {Phys. Rev. C}\ }\textbf {\bibinfo {volume}
  {89}}~(\bibinfo {number} {1}),\ \bibinfo {pages} {015502}},\ \Eprint
  {http://arxiv.org/abs/1306.6106} {arXiv:1306.6106 [nucl-ex]} \BibitemShut
  {NoStop}%
\bibitem [{\citenamefont {Aprile}\ \emph {et~al.}(2022)\citenamefont {Aprile}
  \emph {et~al.}}]{XENON:2022evz}%
  \BibitemOpen
  \bibfield  {author} {\bibinfo {author} {\bibnamefont {Aprile}, \bibfnamefont
  {E}},  \emph {et~al.} (\bibinfo {collaboration} {XENON})} (\bibinfo {year}
  {2022}),\ \bibfield  {title} {\enquote {\bibinfo {title} {{Double-Weak Decays
  of $^{124}$Xe and $^{136}$Xe in the XENON1T and XENONnT Experiments}},}\
  }\href {\doibase 10.1103/PhysRevC.106.024328} {\bibfield  {journal} {\bibinfo
   {journal} {Phys. Rev. C}\ }\textbf {\bibinfo {volume} {106}}~(\bibinfo
  {number} {2}),\ \bibinfo {pages} {024328}},\ \Eprint
  {http://arxiv.org/abs/2205.04158} {arXiv:2205.04158 [hep-ex]} \BibitemShut
  {NoStop}%
\bibitem [{\citenamefont {Ball}\ \emph {et~al.}(2000)\citenamefont {Ball},
  \citenamefont {Harris},\ and\ \citenamefont {McFarland}}]{Ball:2000qd}%
  \BibitemOpen
  \bibfield  {author} {\bibinfo {author} {\bibnamefont {Ball}, \bibfnamefont
  {Richard~D}}, \bibinfo {author} {\bibfnamefont {Deborah~A.}\ \bibnamefont
  {Harris}}, \ and\ \bibinfo {author} {\bibfnamefont {Kevin~Scott}\
  \bibnamefont {McFarland}}} (\bibinfo {year} {2000}),\ \bibfield  {title}
  {\enquote {\bibinfo {title} {{Flavor decomposition of nucleon structure at a
  neutrino factory}},}\ }in\ \href@noop {} {\emph {\bibinfo {booktitle}
  {{NuFACT'00: International Workshop on Muon Storage Ring for a Neutrino
  Factory}}}},\ \Eprint {http://arxiv.org/abs/hep-ph/0009223}
  {arXiv:hep-ph/0009223} \BibitemShut {NoStop}%
\bibitem [{\citenamefont {Bardeen}(1995)}]{Bardeen:1995kv}%
  \BibitemOpen
  \bibfield  {author} {\bibinfo {author} {\bibnamefont {Bardeen}, \bibfnamefont
  {William~A}}} (\bibinfo {year} {1995}),\ \bibfield  {title} {\enquote
  {\bibinfo {title} {{On naturalness in the standard model}},}\ }in\ \href@noop
  {} {\emph {\bibinfo {booktitle} {{Ontake Summer Institute on Particle
  Physics}}}},\ \bibinfo {note} {{FERMILAB-CONF-95-391-T}}\BibitemShut
  {NoStop}%
\bibitem [{\citenamefont {Bjorken}(1983)}]{Bjorken:1982iw}%
  \BibitemOpen
  \bibfield  {author} {\bibinfo {author} {\bibnamefont {Bjorken}, \bibfnamefont
  {J~D}}} (\bibinfo {year} {1983}),\ \enquote {\bibinfo {title} {{A Thousand
  TeV in the Center-of-Mass: Introduction to High-Energy Storage Rings}},}\ in\
  \href {\doibase 10.1007/978-1-4613-3745-4_6} {\emph {\bibinfo {booktitle}
  {{Techniques and Concepts of High-Energy Physics: Proceedings, 2nd NATO
  Advanced Study Institute, Lake George, New York, July 1-12, 1982}}}},\
  Vol.~\bibinfo {volume} {99},\ \bibinfo {editor} {edited by\ \bibinfo {editor}
  {\bibfnamefont {T.}~\bibnamefont {Ferbel}}},\ pp.\ \bibinfo {pages}
  {233--300},\ \bibinfo {note} {{FERMILAB-CONF-82-055-THY},
  \url{https://link.springer.com/content/pdf/10.1007/978-1-4613-3745-4.pdf}}\BibitemShut
  {NoStop}%
\bibitem [{\citenamefont {de~Blas}(2025)}]{deBlas:2944678}%
  \BibitemOpen
  \bibfield  {author} {\bibinfo {author} {\bibnamefont {de~Blas}, \bibfnamefont
  {Jorge, et~al}}} (\bibinfo {year} {2025}),\ \href {\doibase
  10.17181/CERN.35CH.2O2P} {\emph {\bibinfo {title} {{Physics Briefing Book:
  Input for the 2026 update of the European Strategy for Particle Physics}}}},\
  \bibinfo {type} {Tech. Rep.}\ (\bibinfo {address} {Geneva})\ \bibinfo {note}
  {\url{https://cds.cern.ch/record/2944678}}\BibitemShut {NoStop}%
\bibitem [{\citenamefont {Bloise}\ \emph {et~al.}(2025)\citenamefont {Bloise}
  \emph {et~al.}}]{Bloise:2025buj}%
  \BibitemOpen
  \bibfield  {author} {\bibinfo {author} {\bibnamefont {Bloise}, \bibfnamefont
  {C}},  \emph {et~al.}} (\bibinfo {year} {2025}),\ \bibfield  {title}
  {\enquote {\bibinfo {title} {{Sustainability Assessment of Future
  Accelerators}},}\ }\href@noop {} {\ }\Eprint
  {http://arxiv.org/abs/2509.11705} {arXiv:2509.11705 [physics.acc-ph]}
  \BibitemShut {NoStop}%
\bibitem [{\citenamefont {Bloom}\ and\ \citenamefont
  {Boisvert}(2025)}]{Bloom:2024ixe}%
  \BibitemOpen
  \bibfield  {author} {\bibinfo {author} {\bibnamefont {Bloom}, \bibfnamefont
  {Kenneth}}, \ and\ \bibinfo {author} {\bibfnamefont {V{\'e}ronique}\
  \bibnamefont {Boisvert}}} (\bibinfo {year} {2025}),\ \bibfield  {title}
  {\enquote {\bibinfo {title} {{Sustainability and Carbon Emissions of Future
  Accelerators}},}\ }\href {\doibase 10.1146/annurev-nucl-121423-100906}
  {\bibfield  {journal} {\bibinfo  {journal} {Annual Review of Nuclear and
  Particle Science}\ }\textbf {\bibinfo {volume} {75}}~(\bibinfo {number}
  {Volume 75, 2025}),\ \bibinfo {pages} {35--55}},\ \Eprint
  {http://arxiv.org/abs/2411.03473} {arXiv:2411.03473 [hep-ex]} \BibitemShut
  {NoStop}%
\bibitem [{\citenamefont {Bloom}\ \emph {et~al.}(2022)\citenamefont {Bloom},
  \citenamefont {Boisvert}, \citenamefont {Britzger}, \citenamefont {Buuck},
  \citenamefont {Eichhorn}, \citenamefont {Headley}, \citenamefont
  {Lohwasser},\ and\ \citenamefont {Merkel}}]{Bloom:2022gux}%
  \BibitemOpen
  \bibfield  {author} {\bibinfo {author} {\bibnamefont {Bloom}, \bibfnamefont
  {Kenneth}}, \bibinfo {author} {\bibfnamefont {Veronique}\ \bibnamefont
  {Boisvert}}, \bibinfo {author} {\bibfnamefont {Daniel}\ \bibnamefont
  {Britzger}}, \bibinfo {author} {\bibfnamefont {Micah}\ \bibnamefont {Buuck}},
  \bibinfo {author} {\bibfnamefont {Astrid}\ \bibnamefont {Eichhorn}}, \bibinfo
  {author} {\bibfnamefont {Michael}\ \bibnamefont {Headley}}, \bibinfo {author}
  {\bibfnamefont {Kristin}\ \bibnamefont {Lohwasser}}, \ and\ \bibinfo {author}
  {\bibfnamefont {Petra}\ \bibnamefont {Merkel}}} (\bibinfo {year} {2022}),\
  \bibfield  {title} {\enquote {\bibinfo {title} {{Climate impacts of particle
  physics}},}\ }in\ \href@noop {} {\emph {\bibinfo {booktitle} {{Snowmass
  2021}}}},\ \Eprint {http://arxiv.org/abs/2203.12389} {arXiv:2203.12389
  [physics.soc-ph]} \BibitemShut {NoStop}%
\bibitem [{\citenamefont {Bogacz}\ \emph {et~al.}(2022)\citenamefont {Bogacz}
  \emph {et~al.}}]{Bogacz:2022xsj}%
  \BibitemOpen
  \bibfield  {author} {\bibinfo {author} {\bibnamefont {Bogacz}, \bibfnamefont
  {Alex}},  \emph {et~al.}} (\bibinfo {year} {2022}),\ \bibfield  {title}
  {\enquote {\bibinfo {title} {{The Physics Case for a Neutrino Factory}},}\
  }in\ \href@noop {} {\emph {\bibinfo {booktitle} {{Snowmass 2021}}}},\ \Eprint
  {http://arxiv.org/abs/2203.08094} {arXiv:2203.08094 [hep-ph]} \BibitemShut
  {NoStop}%
\bibitem [{\citenamefont {Butler}\ \emph {et~al.}(2023)\citenamefont {Butler}
  \emph {et~al.}}]{Butler:2023glv}%
  \BibitemOpen
  \bibinfo {editor} {\bibnamefont {Butler}, \bibfnamefont {Joel~N}},  \emph
  {et~al.},\ Eds. (\bibinfo {year} {2023}),\ \href {\doibase 10.2172/1922503}
  {\emph {\bibinfo {title} {{Report of the 2021 U.S. Community Study on the
  Future of Particle Physics (Snowmass 2021)}}}},\ \bibinfo {note}
  {\href{https://www.slac.stanford.edu/econf/C210711/SnowmassBook.pdf}{eConf
  C210711}}\BibitemShut {NoStop}%
\bibitem [{\citenamefont {Cahn}(1996)}]{Cahn:1996ag}%
  \BibitemOpen
  \bibfield  {author} {\bibinfo {author} {\bibnamefont {Cahn}, \bibfnamefont
  {Robert~N}}} (\bibinfo {year} {1996}),\ \bibfield  {title} {\enquote
  {\bibinfo {title} {{The Eighteen arbitrary parameters of the standard model
  in your everyday life}},}\ }\href {\doibase 10.1103/RevModPhys.68.951}
  {\bibfield  {journal} {\bibinfo  {journal} {Rev. Mod. Phys.}\ }\textbf
  {\bibinfo {volume} {68}},\ \bibinfo {pages} {951--960}}\BibitemShut {NoStop}%
\bibitem [{\citenamefont {Campana}\ \emph {et~al.}(2016)\citenamefont
  {Campana}, \citenamefont {Klute},\ and\ \citenamefont {Wells}}]{Campana2016}%
  \BibitemOpen
  \bibfield  {author} {\bibinfo {author} {\bibnamefont {Campana}, \bibfnamefont
  {P}}, \bibinfo {author} {\bibfnamefont {M.}~\bibnamefont {Klute}}, \ and\
  \bibinfo {author} {\bibfnamefont {P.~S.}\ \bibnamefont {Wells}}} (\bibinfo
  {year} {2016}),\ \bibfield  {title} {\enquote {\bibinfo {title} {Physics
  goals and experimental challenges of the proton--proton high-luminosity
  operation of the {LHC}},}\ }\href {\doibase
  10.1146/annurev-nucl-102115-044812} {\bibfield  {journal} {\bibinfo
  {journal} {Annual Review of Nuclear and Particle Science}\ }\textbf {\bibinfo
  {volume} {66}}~(\bibinfo {number} {1}),\ \bibinfo {pages} {273--295}},\
  \bibinfo {note} {{Also see the work carried out by the International Linear
  Collider Detector R\&D Groups,
  \url{http://www.linearcollider.org/P-D/Detector-R-D-groups}}}\BibitemShut
  {NoStop}%
\bibitem [{\citenamefont {Chatrchyan}\ \emph {et~al.}(2012)\citenamefont
  {Chatrchyan} \emph {et~al.}}]{CMS:2012qbp}%
  \BibitemOpen
  \bibfield  {author} {\bibinfo {author} {\bibnamefont {Chatrchyan},
  \bibfnamefont {Serguei}},  \emph {et~al.} (\bibinfo {collaboration} {CMS})}
  (\bibinfo {year} {2012}),\ \bibfield  {title} {\enquote {\bibinfo {title}
  {{Observation of a New Boson at a Mass of 125 GeV with the CMS Experiment at
  the LHC}},}\ }\href {\doibase 10.1016/j.physletb.2012.08.021} {\bibfield
  {journal} {\bibinfo  {journal} {Phys. Lett. B}\ }\textbf {\bibinfo {volume}
  {716}},\ \bibinfo {pages} {30--61}},\ \Eprint
  {http://arxiv.org/abs/1207.7235} {arXiv:1207.7235 [hep-ex]} \BibitemShut
  {NoStop}%
\bibitem [{\citenamefont {{CMS~Collaboration}}(2025)}]{CMS:2025jwz}%
  \BibitemOpen
  \bibfield  {author} {\bibinfo {author} {\bibnamefont {{CMS~Collaboration}},}}
  (\bibinfo {year} {2025}),\ \bibfield  {title} {\enquote {\bibinfo {title}
  {{Combined measurements and interpretations of Higgs boson production and
  decay at $\sqrt s$=13 TeV}},}\ }\href@noop {} {\ }\bibinfo {note}
  {CMS-PAS-HIG-21-018, \url{https://cds.cern.ch/record/2929999}}\BibitemShut
  {NoStop}%
\bibitem [{\citenamefont {Cole}\ \emph {et~al.}(1968)\citenamefont {Cole},
  \citenamefont {Goldwasser},\ and\ \citenamefont {Wilson}}]{Cole:1968wx}%
  \BibitemOpen
  \bibfield  {author} {\bibinfo {author} {\bibnamefont {Cole}, \bibfnamefont
  {Frank~T}}, \bibinfo {author} {\bibfnamefont {Edwin~L.}\ \bibnamefont
  {Goldwasser}}, \ and\ \bibinfo {author} {\bibfnamefont {Robert~Rathbun}\
  \bibnamefont {Wilson}}} (\bibinfo {year} {1968}),\ \href@noop {} {\emph
  {\bibinfo {title} {{National Accelerator Laboratory Design Report}}}},\
  \bibinfo {type} {Tech. Rep.}\ (\bibinfo  {institution} {{Fermilab}})\
  \bibinfo {note}
  {\href{http://inspirehep.net/record/53213/files/fermilab-design-1968-01.pdf}{FERMILAB-DESIGN-1968-01},
  \S2.2}\BibitemShut {NoStop}%
\bibitem [{\citenamefont {Dawson}\ \emph {et~al.}(2019)\citenamefont {Dawson},
  \citenamefont {Englert},\ and\ \citenamefont {Plehn}}]{Dawson:2018dcd}%
  \BibitemOpen
  \bibfield  {author} {\bibinfo {author} {\bibnamefont {Dawson}, \bibfnamefont
  {Sally}}, \bibinfo {author} {\bibfnamefont {Christoph}\ \bibnamefont
  {Englert}}, \ and\ \bibinfo {author} {\bibfnamefont {Tilman}\ \bibnamefont
  {Plehn}}} (\bibinfo {year} {2019}),\ \bibfield  {title} {\enquote {\bibinfo
  {title} {{Higgs Physics: It ain't over till it's over}},}\ }\href {\doibase
  10.1016/j.physrep.2019.05.001} {\bibfield  {journal} {\bibinfo  {journal}
  {Phys. Rept.}\ }\textbf {\bibinfo {volume} {816}},\ \bibinfo {pages}
  {1--85}},\ \Eprint {http://arxiv.org/abs/1808.01324} {arXiv:1808.01324
  [hep-ph]} \BibitemShut {NoStop}%
\bibitem [{\citenamefont {Denton}\ and\ \citenamefont
  {Gehrlein}(2025)}]{Denton:2024glz}%
  \BibitemOpen
  \bibfield  {author} {\bibinfo {author} {\bibnamefont {Denton}, \bibfnamefont
  {Peter~B}}, \ and\ \bibinfo {author} {\bibfnamefont {Julia}\ \bibnamefont
  {Gehrlein}}} (\bibinfo {year} {2025}),\ \bibfield  {title} {\enquote
  {\bibinfo {title} {{A modern look at the oscillation physics case for a
  neutrino factory}},}\ }\href {\doibase 10.1016/j.nuclphysb.2025.116818}
  {\bibfield  {journal} {\bibinfo  {journal} {Nucl. Phys. B}\ }\textbf
  {\bibinfo {volume} {1012}},\ \bibinfo {pages} {116818}},\ \Eprint
  {http://arxiv.org/abs/2407.02572} {arXiv:2407.02572 [hep-ph]} \BibitemShut
  {NoStop}%
\bibitem [{\citenamefont {Dicke}\ \emph {et~al.}(1965)\citenamefont {Dicke},
  \citenamefont {Peebles}, \citenamefont {Roll},\ and\ \citenamefont
  {Wilkinson}}]{Dicke:1965zz}%
  \BibitemOpen
  \bibfield  {author} {\bibinfo {author} {\bibnamefont {Dicke}, \bibfnamefont
  {R~H}}, \bibinfo {author} {\bibfnamefont {P.~J.~E.}\ \bibnamefont {Peebles}},
  \bibinfo {author} {\bibfnamefont {P.~G.}\ \bibnamefont {Roll}}, \ and\
  \bibinfo {author} {\bibfnamefont {D.~T.}\ \bibnamefont {Wilkinson}}}
  (\bibinfo {year} {1965}),\ \bibfield  {title} {\enquote {\bibinfo {title}
  {{Cosmic Black-Body Radiation}},}\ }\href {\doibase 10.1086/148306}
  {\bibfield  {journal} {\bibinfo  {journal} {Astrophys. J.}\ }\textbf
  {\bibinfo {volume} {142}},\ \bibinfo {pages} {414--419}}\BibitemShut
  {NoStop}%
\bibitem [{\citenamefont {Dine}(2015)}]{Dine:2015xga}%
  \BibitemOpen
  \bibfield  {author} {\bibinfo {author} {\bibnamefont {Dine}, \bibfnamefont
  {Michael}}} (\bibinfo {year} {2015}),\ \bibfield  {title} {\enquote {\bibinfo
  {title} {{Naturalness Under Stress}},}\ }\href {\doibase
  10.1146/annurev-nucl-102014-022053} {\bibfield  {journal} {\bibinfo
  {journal} {Ann. Rev. Nucl. Part. Sci.}\ }\textbf {\bibinfo {volume} {65}},\
  \bibinfo {pages} {43--62}},\ \Eprint {http://arxiv.org/abs/1501.01035}
  {arXiv:1501.01035 [hep-ph]} \BibitemShut {NoStop}%
\bibitem [{\citenamefont {Ellis}\ and\ \citenamefont
  {Sakurai}(2016)}]{Ellis:2016ast}%
  \BibitemOpen
  \bibfield  {author} {\bibinfo {author} {\bibnamefont {Ellis}, \bibfnamefont
  {John}}, \ and\ \bibinfo {author} {\bibfnamefont {Kazuki}\ \bibnamefont
  {Sakurai}}} (\bibinfo {year} {2016}),\ \bibfield  {title} {\enquote {\bibinfo
  {title} {{Search for Sphalerons in Proton-Proton Collisions}},}\ }\href
  {\doibase 10.1007/JHEP04(2016)086} {\bibfield  {journal} {\bibinfo  {journal}
  {JHEP}\ }\textbf {\bibinfo {volume} {04}},\ \bibinfo {pages} {086}},\ \Eprint
  {http://arxiv.org/abs/1601.03654} {arXiv:1601.03654 [hep-ph]} \BibitemShut
  {NoStop}%
\bibitem [{\citenamefont {Elvira}\ \emph {et~al.}(2022)\citenamefont {Elvira}
  \emph {et~al.}}]{Elvira:2022wyn}%
  \BibitemOpen
  \bibfield  {author} {\bibinfo {author} {\bibnamefont {Elvira}, \bibfnamefont
  {V~Daniel}},  \emph {et~al.}} (\bibinfo {year} {2022}),\ \bibfield  {title}
  {\enquote {\bibinfo {title} {{The Future of High Energy Physics Software and
  Computing}},}\ }in\ \href {\doibase 10.2172/1898754} {\emph {\bibinfo
  {booktitle} {{Snowmass 2021}}}},\ \Eprint {http://arxiv.org/abs/2210.05822}
  {arXiv:2210.05822 [hep-ex]} \BibitemShut {NoStop}%
\bibitem [{\citenamefont {Frieman}\ \emph {et~al.}(2008)\citenamefont
  {Frieman}, \citenamefont {Turner},\ and\ \citenamefont
  {Huterer}}]{Frieman:2008sn}%
  \BibitemOpen
  \bibfield  {author} {\bibinfo {author} {\bibnamefont {Frieman}, \bibfnamefont
  {Joshua}}, \bibinfo {author} {\bibfnamefont {Michael}\ \bibnamefont
  {Turner}}, \ and\ \bibinfo {author} {\bibfnamefont {Dragan}\ \bibnamefont
  {Huterer}}} (\bibinfo {year} {2008}),\ \bibfield  {title} {\enquote {\bibinfo
  {title} {{Dark Energy and the Accelerating Universe}},}\ }\href {\doibase
  10.1146/annurev.astro.46.060407.145243} {\bibfield  {journal} {\bibinfo
  {journal} {Ann. Rev. Astron. Astrophys.}\ }\textbf {\bibinfo {volume} {46}},\
  \bibinfo {pages} {385--432}},\ \Eprint {http://arxiv.org/abs/0803.0982}
  {arXiv:0803.0982 [astro-ph]} \BibitemShut {NoStop}%
\bibitem [{\citenamefont {Froidevaux}\ and\ \citenamefont
  {Sphicas}(2006)}]{Froidevaux2006}%
  \BibitemOpen
  \bibfield  {author} {\bibinfo {author} {\bibnamefont {Froidevaux},
  \bibfnamefont {Daniel}}, \ and\ \bibinfo {author} {\bibfnamefont {Paris}\
  \bibnamefont {Sphicas}}} (\bibinfo {year} {2006}),\ \bibfield  {title}
  {\enquote {\bibinfo {title} {General-purpose detectors for the large hadron
  collider},}\ }\href {\doibase 10.1146/annurev.nucl.54.070103.181209}
  {\bibfield  {journal} {\bibinfo  {journal} {Annual Review of Nuclear and
  Particle Science}\ }\textbf {\bibinfo {volume} {56}}~(\bibinfo {number}
  {1}),\ \bibinfo {pages} {375--440}}\BibitemShut {NoStop}%
\bibitem [{\citenamefont {Giudice}(2019)}]{Giudice:2017pzm}%
  \BibitemOpen
  \bibfield  {author} {\bibinfo {author} {\bibnamefont {Giudice}, \bibfnamefont
  {Gian~Francesco}}} (\bibinfo {year} {2019}),\ \enquote {\bibinfo {title}
  {{The Dawn of the Post-Naturalness Era}},}\ in\ \href {\doibase
  10.1142/9789813238053_0013} {\emph {\bibinfo {booktitle} {{From My Vast
  Repertoire ...}: {Guido Altarelli's Legacy}}}},\ \bibinfo {editor} {edited
  by\ \bibinfo {editor} {\bibfnamefont {Aharon}\ \bibnamefont {Levy}}, \bibinfo
  {editor} {\bibfnamefont {Stefano}\ \bibnamefont {Forte}}, \ and\ \bibinfo
  {editor} {\bibfnamefont {Giovanni}\ \bibnamefont {Ridolfi}}},\ pp.\ \bibinfo
  {pages} {267--292},\ \Eprint {http://arxiv.org/abs/1710.07663}
  {arXiv:1710.07663 [physics.hist-ph]} \BibitemShut {NoStop}%
\bibitem [{\citenamefont {de~Gouv{\^e}a}\ and\ \citenamefont
  {Thompson}(2025)}]{deGouvea:2025zfq}%
  \BibitemOpen
  \bibfield  {author} {\bibinfo {author} {\bibnamefont {de~Gouv{\^e}a},
  \bibfnamefont {Andr{\'e}}}, \ and\ \bibinfo {author} {\bibfnamefont {Adrian}\
  \bibnamefont {Thompson}}} (\bibinfo {year} {2025}),\ \bibfield  {title}
  {\enquote {\bibinfo {title} {{Electroweak Observables in Neutrino-Electron
  Scattering from a Muon Storage Ring}},}\ }\href@noop {} {\ }\Eprint
  {http://arxiv.org/abs/2505.00152} {arXiv:2505.00152 [hep-ph]} \BibitemShut
  {NoStop}%
\bibitem [{\citenamefont {Grefsrud}\ \emph {et~al.}(2024)\citenamefont
  {Grefsrud}, \citenamefont {Buanes}, \citenamefont {Koutroulis}, \citenamefont
  {Lipniacka}, \citenamefont {Mase{\l}ek}, \citenamefont {Papaefstathiou},
  \citenamefont {Sakurai}, \citenamefont {Sjursen},\ and\ \citenamefont
  {Slazyk}}]{Grefsrud_2024}%
  \BibitemOpen
  \bibfield  {author} {\bibinfo {author} {\bibnamefont {Grefsrud},
  \bibfnamefont {Aurora~Singstad}}, \bibinfo {author} {\bibfnamefont {Trygve}\
  \bibnamefont {Buanes}}, \bibinfo {author} {\bibfnamefont {Fotis}\
  \bibnamefont {Koutroulis}}, \bibinfo {author} {\bibfnamefont {Anna}\
  \bibnamefont {Lipniacka}}, \bibinfo {author} {\bibfnamefont {Rafa\l}\
  \bibnamefont {Mase{\l}ek}}, \bibinfo {author} {\bibfnamefont {Andreas}\
  \bibnamefont {Papaefstathiou}}, \bibinfo {author} {\bibfnamefont {Kazuki}\
  \bibnamefont {Sakurai}}, \bibinfo {author} {\bibfnamefont {Therese~B.}\
  \bibnamefont {Sjursen}}, \ and\ \bibinfo {author} {\bibfnamefont {Igor}\
  \bibnamefont {Slazyk}}} (\bibinfo {year} {2024}),\ \bibfield  {title}
  {\enquote {\bibinfo {title} {Machine learning classification of sphalerons
  and black holes at the {LHC}},}\ }\href {\doibase
  10.1140/epjc/s10052-024-12790-x} {\bibfield  {journal} {\bibinfo  {journal}
  {The European Physical Journal C}\ }\textbf {\bibinfo {volume}
  {84}}~(\bibinfo {number} {4}),\ 10.1140/epjc/s10052-024-12790-x}\BibitemShut
  {NoStop}%
\bibitem [{\citenamefont {Jaffe}\ and\ \citenamefont {Witten}(2000)}]{Massgap}%
  \BibitemOpen
  \bibfield  {author} {\bibinfo {author} {\bibnamefont {Jaffe}, \bibfnamefont
  {Arthur}}, \ and\ \bibinfo {author} {\bibfnamefont {Edward}\ \bibnamefont
  {Witten}}} (\bibinfo {year} {2000}),\ \bibfield  {title} {\enquote {\bibinfo
  {title} {{Quantum Yang--Mills Theory}},}\ }\href@noop {} {\ }\bibinfo {note}
  {\href{https://www.claymath.org/wp-content/uploads/2022/06/yangmills.pdf}{www.claymath.org/wp-content/uploads/2022/06/yangmills.pdf}}\BibitemShut
  {NoStop}%
\bibitem [{\citenamefont {Kajita}(2016)}]{Kajita:2016cak}%
  \BibitemOpen
  \bibfield  {author} {\bibinfo {author} {\bibnamefont {Kajita}, \bibfnamefont
  {Takaaki}}} (\bibinfo {year} {2016}),\ \bibfield  {title} {\enquote {\bibinfo
  {title} {{Nobel Lecture: Discovery of atmospheric neutrino oscillations}},}\
  }\href {\doibase 10.1103/RevModPhys.88.030501} {\bibfield  {journal}
  {\bibinfo  {journal} {Rev. Mod. Phys.}\ }\textbf {\bibinfo {volume}
  {88}}~(\bibinfo {number} {3}),\ \bibinfo {pages} {030501}}\BibitemShut
  {NoStop}%
\bibitem [{\citenamefont {Koppenburg}(2025)}]{PKopp}%
  \BibitemOpen
  \bibfield  {author} {\bibinfo {author} {\bibnamefont {Koppenburg},
  \bibfnamefont {Patrick}}} (\bibinfo {year} {2025}),\ \href@noop {} {\enquote
  {\bibinfo {title} {{New particles discovered at the LHC}},}\ }\bibinfo {note}
  {\href{https://www.koppenburg.ch/particles.html}{www.koppenburg.ch/particles.html}}\BibitemShut
  {NoStop}%
\bibitem [{\citenamefont {Kostelecky}\ and\ \citenamefont
  {Russell}(2011)}]{Kostelecky:2008ts}%
  \BibitemOpen
  \bibfield  {author} {\bibinfo {author} {\bibnamefont {Kostelecky},
  \bibfnamefont {V~Alan}}, \ and\ \bibinfo {author} {\bibfnamefont {Neil}\
  \bibnamefont {Russell}}} (\bibinfo {year} {2011}),\ \bibfield  {title}
  {\enquote {\bibinfo {title} {{Data Tables for Lorentz and CPT Violation}},}\
  }\href {\doibase 10.1103/RevModPhys.83.11} {\bibfield  {journal} {\bibinfo
  {journal} {Rev. Mod. Phys.}\ }\textbf {\bibinfo {volume} {83}},\ \bibinfo
  {pages} {11--31}},\ \Eprint {http://arxiv.org/abs/0801.0287} {arXiv:0801.0287
  [hep-ph]} \BibitemShut {NoStop}%
\bibitem [{\citenamefont {Lahav}\ and\ \citenamefont
  {Liddle}(2024)}]{Lahav:2024npe}%
  \BibitemOpen
  \bibfield  {author} {\bibinfo {author} {\bibnamefont {Lahav}, \bibfnamefont
  {Ofer}}, \ and\ \bibinfo {author} {\bibfnamefont {Andrew~R.}\ \bibnamefont
  {Liddle}}} (\bibinfo {year} {2024}),\ \bibfield  {title} {\enquote {\bibinfo
  {title} {{The Cosmological Parameters (2023)}},}\ }\href@noop {} {\ }\bibinfo
  {note} {In \cite{ParticleDataGroup:2024cfk}, \S25},\ \Eprint
  {http://arxiv.org/abs/2403.15526} {arXiv:2403.15526 [astro-ph.CO]}
  \BibitemShut {NoStop}%
\bibitem [{\citenamefont {Lee}\ \emph {et~al.}(1977)\citenamefont {Lee},
  \citenamefont {Quigg},\ and\ \citenamefont {Thacker}}]{Lee:1977eg}%
  \BibitemOpen
  \bibfield  {author} {\bibinfo {author} {\bibnamefont {Lee}, \bibfnamefont
  {Benjamin~W}}, \bibinfo {author} {\bibfnamefont {C.}~\bibnamefont {Quigg}}, \
  and\ \bibinfo {author} {\bibfnamefont {H.~B.}\ \bibnamefont {Thacker}}}
  (\bibinfo {year} {1977}),\ \bibfield  {title} {\enquote {\bibinfo {title}
  {{Weak Interactions at Very High-Energies: The Role of the Higgs Boson
  Mass}},}\ }\href {\doibase 10.1103/PhysRevD.16.1519} {\bibfield  {journal}
  {\bibinfo  {journal} {Phys. Rev. D}\ }\textbf {\bibinfo {volume} {16}},\
  \bibinfo {pages} {1519}}\BibitemShut {NoStop}%
\bibitem [{\citenamefont {Liberati}(2013)}]{Liberati:2013xla}%
  \BibitemOpen
  \bibfield  {author} {\bibinfo {author} {\bibnamefont {Liberati},
  \bibfnamefont {Stefano}}} (\bibinfo {year} {2013}),\ \bibfield  {title}
  {\enquote {\bibinfo {title} {{Tests of Lorentz invariance: a 2013 update}},}\
  }\href {\doibase 10.1088/0264-9381/30/13/133001} {\bibfield  {journal}
  {\bibinfo  {journal} {Class. Quant. Grav.}\ }\textbf {\bibinfo {volume}
  {30}},\ \bibinfo {pages} {133001}},\ \Eprint {http://arxiv.org/abs/1304.5795}
  {arXiv:1304.5795 [gr-qc]} \BibitemShut {NoStop}%
\bibitem [{\citenamefont {McDonald}(2016)}]{McDonald:2016ixn}%
  \BibitemOpen
  \bibfield  {author} {\bibinfo {author} {\bibnamefont {McDonald},
  \bibfnamefont {Arthur~B}}} (\bibinfo {year} {2016}),\ \bibfield  {title}
  {\enquote {\bibinfo {title} {{Nobel Lecture: The Sudbury Neutrino
  Observatory: Observation of flavor change for solar neutrinos}},}\ }\href
  {\doibase 10.1103/RevModPhys.88.030502} {\bibfield  {journal} {\bibinfo
  {journal} {Rev. Mod. Phys.}\ }\textbf {\bibinfo {volume} {88}}~(\bibinfo
  {number} {3}),\ \bibinfo {pages} {030502}}\BibitemShut {NoStop}%
\bibitem [{\citenamefont {Morrison}\ \emph {et~al.}(1982)\citenamefont
  {Morrison}, \citenamefont {Morrison},\ and\ \citenamefont {{The Office of
  Charles and Ray Eames}}}]{Eames:POT}%
  \BibitemOpen
  \bibfield  {author} {\bibinfo {author} {\bibnamefont {Morrison},
  \bibfnamefont {Philip}}, \bibinfo {author} {\bibfnamefont {Phylis}\
  \bibnamefont {Morrison}}, \ and\ \bibinfo {author} {\bibnamefont {{The Office
  of Charles and Ray Eames}}}} (\bibinfo {year} {1982}),\ \href@noop {} {\emph
  {\bibinfo {title} {{Powers of Ten}}}}\ (\bibinfo  {publisher} {Scientific
  American Books},\ \bibinfo {address} {Redding, CT})\ \bibinfo {note} {{ISBN:
  9780716714095. The book was based on a celebrated Eames film.}}\BibitemShut
  {Stop}%
\bibitem [{\citenamefont {Navas}\ \emph {et~al.}(2024)\citenamefont {Navas}
  \emph {et~al.}}]{ParticleDataGroup:2024cfk}%
  \BibitemOpen
  \bibfield  {author} {\bibinfo {author} {\bibnamefont {Navas}, \bibfnamefont
  {S}},  \emph {et~al.} (\bibinfo {collaboration} {Particle Data Group})}
  (\bibinfo {year} {2024}),\ \bibfield  {title} {\enquote {\bibinfo {title}
  {{Review of Particle Physics}},}\ }\href {\doibase
  10.1103/PhysRevD.110.030001} {\bibfield  {journal} {\bibinfo  {journal}
  {Phys. Rev. D}\ }\textbf {\bibinfo {volume} {110}}~(\bibinfo {number} {3}),\
  \bibinfo {pages} {030001}},\ \bibinfo {note} {and 2025 update:
  \href{https://pdg.lbl.gov/2025/}{pdg.lbl.gov/2025/}}\BibitemShut {NoStop}%
\bibitem [{\citenamefont {Olive}\ and\ \citenamefont
  {Peacock}(2024)}]{OlivePeacock}%
  \BibitemOpen
  \bibfield  {author} {\bibinfo {author} {\bibnamefont {Olive}, \bibfnamefont
  {K~A}}, \ and\ \bibinfo {author} {\bibfnamefont {J.~A.}\ \bibnamefont
  {Peacock}}} (\bibinfo {year} {2024}),\ \bibfield  {title} {\enquote {\bibinfo
  {title} {{Big-Bang Cosmology}},}\ }\href@noop {} {\ }\bibinfo {note} {{in
  \cite{ParticleDataGroup:2024cfk}, \S22.}}\BibitemShut {Stop}%
\bibitem [{\citenamefont {Pati}\ and\ \citenamefont
  {Salam}(1974)}]{Pati:1974yy}%
  \BibitemOpen
  \bibfield  {author} {\bibinfo {author} {\bibnamefont {Pati}, \bibfnamefont
  {Jogesh~C}}, \ and\ \bibinfo {author} {\bibfnamefont {Abdus}\ \bibnamefont
  {Salam}}} (\bibinfo {year} {1974}),\ \bibfield  {title} {\enquote {\bibinfo
  {title} {{Lepton Number as the Fourth Color}},}\ }\href {\doibase
  10.1103/PhysRevD.10.275} {\bibfield  {journal} {\bibinfo  {journal} {Phys.
  Rev. D}\ }\textbf {\bibinfo {volume} {10}},\ \bibinfo {pages} {275--289}},\
  \bibinfo {note} {[Erratum: Phys.Rev.D 11, 703--703 (1975)]}\BibitemShut
  {NoStop}%
\bibitem [{\citenamefont {Payne}(1925)}]{Payne:1925}%
  \BibitemOpen
  \bibfield  {author} {\bibinfo {author} {\bibnamefont {Payne}, \bibfnamefont
  {Cecilia~H}}} (\bibinfo {year} {1925}),\ \href@noop {} {\emph {\bibinfo
  {title} {{Stellar Atmospheres}}}}\ (\bibinfo  {publisher} {Harvard
  Observatory},\ \bibinfo {address} {Cambridge, Massachusetts})\ \bibinfo
  {note}
  {\url{https://articles.adsabs.harvard.edu/pdf/1925HarMo...1.....P}}\BibitemShut
  {NoStop}%
\bibitem [{\citenamefont {Penzias}\ and\ \citenamefont
  {Wilson}(1965)}]{Penzias:1965wn}%
  \BibitemOpen
  \bibfield  {author} {\bibinfo {author} {\bibnamefont {Penzias}, \bibfnamefont
  {Arno~A}}, \ and\ \bibinfo {author} {\bibfnamefont {Robert~Woodrow}\
  \bibnamefont {Wilson}}} (\bibinfo {year} {1965}),\ \bibfield  {title}
  {\enquote {\bibinfo {title} {{A Measurement of excess antenna temperature at
  4080-Mc/s}},}\ }\href {\doibase 10.1086/148307} {\bibfield  {journal}
  {\bibinfo  {journal} {Astrophys. J.}\ }\textbf {\bibinfo {volume} {142}},\
  \bibinfo {pages} {419--421}}\BibitemShut {NoStop}%
\bibitem [{\citenamefont {Perlmutter}\ \emph {et~al.}(1999)\citenamefont
  {Perlmutter} \emph {et~al.}}]{SupernovaCosmologyProject:1998vns}%
  \BibitemOpen
  \bibfield  {author} {\bibinfo {author} {\bibnamefont {Perlmutter},
  \bibfnamefont {{S}}},  \emph {et~al.} (\bibinfo {collaboration} {Supernova
  Cosmology Project})} (\bibinfo {year} {1999}),\ \bibfield  {title} {\enquote
  {\bibinfo {title} {{Measurements of $\Omega$ and $\Lambda$ from 42 High
  Redshift Supernovae}},}\ }\href {\doibase 10.1086/307221} {\bibfield
  {journal} {\bibinfo  {journal} {Astrophys. J.}\ }\textbf {\bibinfo {volume}
  {517}},\ \bibinfo {pages} {565--586}},\ \Eprint
  {http://arxiv.org/abs/astro-ph/9812133} {arXiv:astro-ph/9812133} \BibitemShut
  {NoStop}%
\bibitem [{\citenamefont {Perr\'eaud}\ and\ \citenamefont
  {Klapisch}(2025)}]{VraisChercheurs}%
  \BibitemOpen
  \bibfield  {author} {\bibinfo {author} {\bibnamefont {Perr\'eaud},
  \bibfnamefont {Jean-Luc}}, \ and\ \bibinfo {author} {\bibfnamefont
  {C\'edric}\ \bibnamefont {Klapisch}}} (\bibinfo {year} {2025}),\ \bibfield
  {title} {\enquote {\bibinfo {title} {Les vrais chercheurs ne savent pas ce
  qu'ils cherchent},}\ }\href@noop {} {\ }\bibinfo {note} {{Ce qui me meut
  Motion Picture documentary film}}\BibitemShut {NoStop}%
\bibitem [{\citenamefont {Planck}(1899)}]{PlanckUnits:1899}%
  \BibitemOpen
  \bibfield  {author} {\bibinfo {author} {\bibnamefont {Planck}, \bibfnamefont
  {M}}} (\bibinfo {year} {1899}),\ \bibfield  {title} {\enquote {\bibinfo
  {title} {{\"Uber irreversible Strahlungsvorg\"ange}},}\ }\href@noop {}
  {\bibfield  {journal} {\bibinfo  {journal} {Sitz. K\"oniglich Preu{\ss}ischen
  Akad. Wiss. Berlin}\ }\textbf {\bibinfo {volume} {5}},\ \bibinfo {pages}
  {440--480}},\ \bibinfo {note}
  {\url{https://uni-tuebingen.de/fileadmin/Uni_Tuebingen/Fakultaeten/MathePhysik/Institute/IAP/Forschung/MOettel/Geburt_QM/planck_AnnPhys_306_69_1900.pdf},
  \S26}\BibitemShut {NoStop}%
\bibitem [{\citenamefont {Pomeranchuk}(1958)}]{Pomeranchuk:1958ged}%
  \BibitemOpen
  \bibfield  {author} {\bibinfo {author} {\bibnamefont {Pomeranchuk},
  \bibfnamefont {I~Ia}}} (\bibinfo {year} {1958}),\ \bibfield  {title}
  {\enquote {\bibinfo {title} {{Equality of the Nucleon and Antinucleon Total
  Interaction Cross Section at High Energies}},}\ }\href@noop {} {\bibfield
  {journal} {\bibinfo  {journal} {Sov. Phys. JETP}\ }\textbf {\bibinfo {volume}
  {7}},\ \bibinfo {pages} {499--501}}\BibitemShut {NoStop}%
\bibitem [{\citenamefont {Powell}(1969?)}]{Powell:frag}%
  \BibitemOpen
  \bibfield  {author} {\bibinfo {author} {\bibnamefont {Powell}, \bibfnamefont
  {Cecil}}} (\bibinfo {year} {1969?}),\ \bibfield  {title} {\enquote {\bibinfo
  {title} {Fragments of autobiography},}\ }\href@noop {} {\ }\bibinfo {note}
  {\url{https://www.bristol.ac.uk/physics/media/histories/12-powell.pdf}}\BibitemShut
  {NoStop}%
\bibitem [{\citenamefont {{Quarkonium Working Group}}(2025)}]{QWGExotics}%
  \BibitemOpen
  \bibfield  {author} {\bibinfo {author} {\bibnamefont {{Quarkonium Working
  Group}},}} (\bibinfo {year} {2025}),\ \href@noop {} {\enquote {\bibinfo
  {title} {{Exotics Hub}},}\ }\bibinfo {note}
  {\href{https://https://qwg.ph.nat.tum.de/exoticshub/}{qwg.ph.nat.tum.de/exoticshub/}}\BibitemShut
  {NoStop}%
\bibitem [{\citenamefont {Quigg}(2009)}]{Quigg:2009vq}%
  \BibitemOpen
  \bibfield  {author} {\bibinfo {author} {\bibnamefont {Quigg}, \bibfnamefont
  {Chris}}} (\bibinfo {year} {2009}),\ \bibfield  {title} {\enquote {\bibinfo
  {title} {{Unanswered Questions in the Electroweak Theory}},}\ }\href
  {\doibase 10.1146/annurev.nucl.010909.083126} {\bibfield  {journal} {\bibinfo
   {journal} {Ann. Rev. Nucl. Part. Sci.}\ }\textbf {\bibinfo {volume} {59}},\
  \bibinfo {pages} {505--555}}\BibitemShut {NoStop}%
\bibitem [{\citenamefont {Quigg}(2010)}]{Quigg:2010nn}%
  \BibitemOpen
  \bibfield  {author} {\bibinfo {author} {\bibnamefont {Quigg}, \bibfnamefont
  {Chris}}} (\bibinfo {year} {2010}),\ \bibfield  {title} {\enquote {\bibinfo
  {title} {{Learning to See at the Large Hadron Collider}},}\ }\href@noop {} {\
  }\Eprint {http://arxiv.org/abs/1001.2025} {arXiv:1001.2025 [hep-ph]}
  \BibitemShut {NoStop}%
\bibitem [{\citenamefont {Quigg}(2015)}]{Quigg:2015cfa}%
  \BibitemOpen
  \bibfield  {author} {\bibinfo {author} {\bibnamefont {Quigg}, \bibfnamefont
  {Chris}}} (\bibinfo {year} {2015}),\ \bibfield  {title} {\enquote {\bibinfo
  {title} {{Electroweak Symmetry Breaking in Historical Perspective}},}\ }\href
  {\doibase 10.1146/annurev-nucl-102313-025537} {\bibfield  {journal} {\bibinfo
   {journal} {Ann. Rev. Nucl. Part. Sci.}\ }\textbf {\bibinfo {volume} {65}},\
  \bibinfo {pages} {25--42}},\ \Eprint {http://arxiv.org/abs/1503.01756}
  {arXiv:1503.01756 [hep-ph]} \BibitemShut {NoStop}%
\bibitem [{\citenamefont {Quigg}\ and\ \citenamefont
  {Shrock}(2009)}]{Quigg:2009xr}%
  \BibitemOpen
  \bibfield  {author} {\bibinfo {author} {\bibnamefont {Quigg}, \bibfnamefont
  {Chris}}, \ and\ \bibinfo {author} {\bibfnamefont {Robert}\ \bibnamefont
  {Shrock}}} (\bibinfo {year} {2009}),\ \bibfield  {title} {\enquote {\bibinfo
  {title} {{Gedanken Worlds without Higgs: QCD-Induced Electroweak Symmetry
  Breaking}},}\ }\href {\doibase 10.1103/PhysRevD.79.096002} {\bibfield
  {journal} {\bibinfo  {journal} {Phys. Rev. D}\ }\textbf {\bibinfo {volume}
  {79}},\ \bibinfo {pages} {096002}},\ \Eprint {http://arxiv.org/abs/0901.3958}
  {arXiv:0901.3958 [hep-ph]} \BibitemShut {NoStop}%
\bibitem [{\citenamefont {Riess}\ \emph {et~al.}(1998)\citenamefont {Riess}
  \emph {et~al.}}]{SupernovaSearchTeam:1998fmf}%
  \BibitemOpen
  \bibfield  {author} {\bibinfo {author} {\bibnamefont {Riess}, \bibfnamefont
  {Adam~G}},  \emph {et~al.} (\bibinfo {collaboration} {Supernova Search
  Team})} (\bibinfo {year} {1998}),\ \bibfield  {title} {\enquote {\bibinfo
  {title} {{Observational evidence from supernovae for an accelerating universe
  and a cosmological constant}},}\ }\href {\doibase 10.1086/300499} {\bibfield
  {journal} {\bibinfo  {journal} {Astron. J.}\ }\textbf {\bibinfo {volume}
  {116}},\ \bibinfo {pages} {1009--1038}},\ \Eprint
  {http://arxiv.org/abs/astro-ph/9805201} {arXiv:astro-ph/9805201} \BibitemShut
  {NoStop}%
\bibitem [{\citenamefont {Salam}\ \emph {et~al.}(2022)\citenamefont {Salam},
  \citenamefont {Wang},\ and\ \citenamefont {Zanderighi}}]{Salam:2022izo}%
  \BibitemOpen
  \bibfield  {author} {\bibinfo {author} {\bibnamefont {Salam}, \bibfnamefont
  {Gavin~P}}, \bibinfo {author} {\bibfnamefont {Lian-Tao}\ \bibnamefont
  {Wang}}, \ and\ \bibinfo {author} {\bibfnamefont {Giulia}\ \bibnamefont
  {Zanderighi}}} (\bibinfo {year} {2022}),\ \bibfield  {title} {\enquote
  {\bibinfo {title} {{The Higgs boson turns ten}},}\ }\href {\doibase
  10.1038/s41586-022-04899-4} {\bibfield  {journal} {\bibinfo  {journal}
  {Nature}\ }\textbf {\bibinfo {volume} {607}}~(\bibinfo {number} {7917}),\
  \bibinfo {pages} {41--47}},\ \Eprint {http://arxiv.org/abs/2207.00478}
  {arXiv:2207.00478 [hep-ph]} \BibitemShut {NoStop}%
\bibitem [{\citenamefont {Streater}\ and\ \citenamefont
  {Wightman}(1964)}]{PCTSpin}%
  \BibitemOpen
  \bibfield  {author} {\bibinfo {author} {\bibnamefont {Streater},
  \bibfnamefont {R~F}}, \ and\ \bibinfo {author} {\bibfnamefont {A.~S.}\
  \bibnamefont {Wightman}}} (\bibinfo {year} {1964}),\ \href@noop {} {\emph
  {\bibinfo {title} {{PCT, Spin \& Statistics, and All That}}}}\ (\bibinfo
  {publisher} {{W. A. Benjamin}},\ \bibinfo {address} {New York and
  Amsterdam})\BibitemShut {NoStop}%
\bibitem [{\citenamefont {Susskind}(2003)}]{Susskind:2003kw}%
  \BibitemOpen
  \bibfield  {author} {\bibinfo {author} {\bibnamefont {Susskind},
  \bibfnamefont {Leonard}}} (\bibinfo {year} {2003}),\ \bibfield  {title}
  {\enquote {\bibinfo {title} {{The Anthropic Landscape of String Theory}},}\
  }\href@noop {} {\ }\Eprint {http://arxiv.org/abs/hep-th/0302219}
  {arXiv:hep-th/0302219} \BibitemShut {NoStop}%
\bibitem [{\citenamefont {Tumasyan}\ \emph {et~al.}(2022)\citenamefont
  {Tumasyan} \emph {et~al.}}]{CMS:2022dwd}%
  \BibitemOpen
  \bibfield  {author} {\bibinfo {author} {\bibnamefont {Tumasyan},
  \bibfnamefont {Armen}},  \emph {et~al.} (\bibinfo {collaboration} {CMS})}
  (\bibinfo {year} {2022}),\ \bibfield  {title} {\enquote {\bibinfo {title} {{A
  portrait of the Higgs boson by the CMS experiment ten years after the
  discovery.}}}\ }\href {\doibase 10.1038/s41586-022-04892-x} {\bibfield
  {journal} {\bibinfo  {journal} {Nature}\ }\textbf {\bibinfo {volume}
  {607}}~(\bibinfo {number} {7917}),\ \bibinfo {pages} {60--68}},\ \bibinfo
  {note} {[Erratum: Nature 623, (2023)]},\ \Eprint
  {http://arxiv.org/abs/2207.00043} {arXiv:2207.00043 [hep-ex]} \BibitemShut
  {NoStop}%
\bibitem [{\citenamefont {{US Department of Energy}}(2009)}]{AFAF}%
  \BibitemOpen
  \bibfield  {author} {\bibinfo {author} {\bibnamefont {{US Department of
  Energy}},}} (\bibinfo {year} {2009}),\ \href@noop {} {\enquote {\bibinfo
  {title} {{Accelerators for America's Future}},}\ }\bibinfo {note}
  {{\href{https://science.energy.gov/~/media/hep/pdf/accelerator-rd-stewardship/Report.pdf}{U.S.
  Department of Energy Report,
  \url{http://science.energy.gov/~/media/hep/pdf/accelerator-rd-stewardship/Report.pdf}.
  See also \url{http://www.acceleratorsamerica.org}}}}\BibitemShut {NoStop}%
\bibitem [{\citenamefont {Weinberg}(1977)}]{Weinberg:1977ji}%
  \BibitemOpen
  \bibfield  {author} {\bibinfo {author} {\bibnamefont {Weinberg},
  \bibfnamefont {Steven}}} (\bibinfo {year} {1977}),\ \href@noop {} {\emph
  {\bibinfo {title} {{The First Three Minutes. A Modern View of the Origin of
  the Universe}}}}\ (\bibinfo  {publisher} {W. W. Norton},\ \bibinfo {address}
  {New York})\BibitemShut {NoStop}%
\bibitem [{\citenamefont {Weinberg}(2021)}]{Weinberg:2021exr}%
  \BibitemOpen
  \bibfield  {author} {\bibinfo {author} {\bibnamefont {Weinberg},
  \bibfnamefont {Steven}}} (\bibinfo {year} {2021}),\ \bibfield  {title}
  {\enquote {\bibinfo {title} {{On the Development of Effective Field
  Theory}},}\ }\href {\doibase 10.1140/epjh/s13129-021-00004-x} {\bibfield
  {journal} {\bibinfo  {journal} {Eur. Phys. J. H}\ }\textbf {\bibinfo {volume}
  {46}}~(\bibinfo {number} {1}),\ \bibinfo {pages} {6}},\ \Eprint
  {http://arxiv.org/abs/2101.04241} {arXiv:2101.04241 [hep-th]} \BibitemShut
  {NoStop}%
\bibitem [{\citenamefont {Wheeler}\ and\ \citenamefont
  {Ford}(1998)}]{Wheeler:1998}%
  \BibitemOpen
  \bibfield  {author} {\bibinfo {author} {\bibnamefont {Wheeler}, \bibfnamefont
  {John~Archibald}}, \ and\ \bibinfo {author} {\bibfnamefont {Kenneth}\
  \bibnamefont {Ford}}} (\bibinfo {year} {1998}),\ \href@noop {} {\emph
  {\bibinfo {title} {Geons, Black Holes, and Quantum Foam: A Life in
  Physics}}}\ (\bibinfo  {publisher} {W. W. Norton},\ \bibinfo {address} {New
  York})\BibitemShut {NoStop}%
\bibitem [{\citenamefont {Wilczek}(2005)}]{10.1063/1.2138392}%
  \BibitemOpen
  \bibfield  {author} {\bibinfo {author} {\bibnamefont {Wilczek}, \bibfnamefont
  {Frank}}} (\bibinfo {year} {2005}),\ \bibfield  {title} {\enquote {\bibinfo
  {title} {{On Absolute Units, I: Choices}},}\ }\href {\doibase
  10.1063/1.2138392} {\bibfield  {journal} {\bibinfo  {journal} {Physics
  Today}\ }\textbf {\bibinfo {volume} {58}}~(\bibinfo {number} {10}),\ \bibinfo
  {pages} {12--13}}\BibitemShut {NoStop}%
\bibitem [{\citenamefont {Wilson}(2005)}]{WILSON20053}%
  \BibitemOpen
  \bibfield  {author} {\bibinfo {author} {\bibnamefont {Wilson}, \bibfnamefont
  {K~G}}} (\bibinfo {year} {2005}),\ \bibfield  {title} {\enquote {\bibinfo
  {title} {The origins of lattice gauge theory},}\ }\href {\doibase
  https://doi.org/10.1016/j.nuclphysbps.2004.11.271} {\bibfield  {journal}
  {\bibinfo  {journal} {Nuclear Physics B - Proceedings Supplements}\ }\textbf
  {\bibinfo {volume} {140}},\ \bibinfo {pages} {3--19}},\ \bibinfo {note} {\S5:
  ``The final blunder \ldots''}\BibitemShut {NoStop}%
\end{thebibliography}%

\end{document}